\documentclass[twocolumn,showpacs,preprintnumbers,superscriptaddress,amsmath,amssymb,nofootinbib]{revtex4}
\usepackage{epsf}
\usepackage{graphicx}
\newcommand{\cd}{\makebox[0.08cm]{$\cdot$}}
\newcommand{\bg}[1]{\mbox{\boldmath $#1$}}
\newcommand{\sla}{\not\!}
\newcommand{\slaA}{A\!\!\!/}
\newcommand{\slaCA}{{\cal A}\!\!\!/}
\begin{document}
\preprint{PCCF RI 0608}

\title{Regularization of fermion self-energy and\\
electromagnetic vertex in Yukawa model\\
within light-front dynamics}
\author{V.A.~Karmanov}
\affiliation {Lebedev Physical Institute, Leninsky Prospekt 53,
119991 Moscow, Russia}
\author{J.-F.~Mathiot}
\affiliation {Laboratoire de Physique
Corpusculaire, Universit\'e Blaise-Pascal, \\
CNRS/IN2P3, 24 avenue des Landais, F-63177 Aubi\`ere Cedex,
France}
\author{A.V.~Smirnov}
\affiliation {Lebedev Physical Institute, Leninsky Prospekt 53,
119991 Moscow, Russia}

\bibliographystyle{unsrt}

\begin{abstract}
In light-front dynamics,  the  regularization of amplitudes by
traditional cutoffs imposed on the transverse and longitudinal
components of particle momenta corresponds to restricting the
integration volume by  a non-rotationally invariant domain. The
result depends not only on the size of this domain (i.e., on the
cutoff values), but also on its orientation determined by the
position of the light-front plane. Explicitly covariant
formulation of light front dynamics allows us to parameterize the
latter dependence in a very transparent form. If we decompose the
regularized amplitude  in terms of independent invariant
amplitudes,  extra (non-physical) terms should appear, with spin
structures which explicitly depend on the orientation of the light
front plane. The number of form factors, i.e.,  the coefficients
of this decomposition,  therefore also increases. The spin-1/2
fermion self-energy is determined by three scalar functions,
instead of the two standard ones, while for the elastic
electromagnetic vertex the number of   form factors increases from
two to five.  In the present paper we calculate perturbatively all
these form factors in the Yukawa model. Then we compare the
results obtained in the two following ways: ({\it i}) by using the
light front dynamics graph technique rules directly; ({\it ii}) by
integrating the corresponding Feynman amplitudes in terms of the
light front variables. For each of these methods, we use two types
of regularization: the transverse and longitudinal cutoffs, and
the Pauli-Villars regularization. In the latter case, the
dependence of amplitudes on the light front plane orientation
vanishes completely provided enough Pauli-Villars subtractions are
made.
\end{abstract}
\pacs {11.10.-z, 11.25.Db, 13.40.Gp}

\maketitle

\section{Introduction}
Light-Front Dynamics (LFD) is extensively and successfully applied
to  hadron phenomenology, relativistic few-body systems, and field
theory. For reviews of theoretical developments and applications
see, e.g., Refs.~\cite{cdkm,bpp}. In the LFD framework,
non-perturbative approaches to field theory were developed in
Refs.~\cite{Br_et_al_03,Br_et_al_04,Br_et_al_06} and in
Refs.~\cite{BCKM_02,kms_04,mks_05}.  In spite of some essential
differences, these approaches  proceed from the same starting
point, namely, they approximate the state vector of the system by
a truncated one. The problem is then solved without any
decomposition in powers of the coupling constant. Doing that, one
should carry out the renormalization procedure non-perturbatively.
This is a non-trivial problem which is at the heart of ongoing
intense research.  However, before renormalization, one should
regularize the amplitudes, both in the perturbative and
non-perturbative frameworks. The explicit dependence of these
amplitudes on the cutoffs is not unique. It is determined by the
method of regularization.

It is of utmost importance to understand the origin and the
implications of this dependence if one wants to address the
question of non-perturbative renormalization. As we shall show in
the present article, this is the only way to identify the
structure of the counterterms needed to recover full rotationally
invariant renormalized amplitudes in LFD. The question of the
non-perturbative determination of the counterterms in truncated
Fock space is discussed in Ref.~\cite{mks_05}.

The rules for calculating amplitudes can be derived directly,
either by transforming the standard T-ordering of the $S$-matrix
into the ordering along the Light Front (LF)  time (the LFD graph
technique rules~\cite{cdkm}), or from quantized field theory on
the LF plane~\cite{bpp}. An alternative method consists in
expressing well-defined Feynman amplitudes through  the LF
variables and  integrating over the minus-components of particle
momenta~\cite{lb95,Nico1}.  Such an approach was applied to the
derivation and study of the LF electromagnetic
amplitudes~\cite{frederico,bakker_02} and to the analysis of
different contributions to the electromagnetic current, resulting
from the LF reduction of the Bethe-Salpeter
formalism~\cite{miller_03}.  A study of electroweak transitions of
the spin-1 mesons, based on using the LF plane of general
orientation \cite{cdkm}, was carried out in
Refs.~\cite{bissey,jaus_2003}. The comparison of perturbative
amplitudes obtained from the LFD graph technique rules with those
derived from the Feynman approach is, in general, not trivial, as
shown in Ref.~\cite{bakker_05}.

Concerning  the  regularization of amplitudes in LFD, at least two
important features should be mentioned. First, if  a given
physical process is described by a set of LF diagrams, each
partial LF amplitude usually diverges more strongly than the
Feynman amplitude of the same process. The statement holds true
regardless of the origin of the LF contributions: either from the
rules of LFD, or from the Feynman amplitude. This increases the
sensitivity of the result to the choice of the regularization
procedure and may be a source of the so-called treacherous
points~\cite{jibak1}. Second, the LF variables  (and, hence, the
integration domain with the cutoffs imposed on it) explicitly
depend on the LF plane orientation, which means the loss of
rotational invariance. Because of this extra dependence,  {\em
standard} decompositions of such regularized LF amplitudes into
invariant amplitudes are not valid. The total number of invariant
amplitudes (and the number of form factors which are the
coefficients in this decomposition) increases, as compared  to the
case when the rotational invariance is preserved.  For example,
the LF electromagnetic vertex (EMV) of a spin-1/2 particle is
determined by five form factors rather than by two. In order to
cancel the extra contributions (which depend on the LF plane
orientation) one needs to introduce in the interaction Hamiltonian
new specific counterterms.

This  complication is especially dramatic in non-perturbative
approaches, where the  Fock space truncation is another source of
the rotational symmetry violation.  A given Feynman  diagram  may
generate a few time-ordered ones  with intermediate states
containing different number  of particles. When they are
truncated, the rotational invariance is lost even for invariant
cutoffs.  The interlacing of the two sources of the violation of
rotational symmetry makes non-perturbative analysis of the
counterterm structure extremely involved. One should therefore
separate to a maximal extent the problems  coming from the
regularization procedure and  from the  Fock space truncation.
That can be effectively done in the explicitly covariant
formulation of LFD within  the perturbative framework in a given
order in the coupling constant.

In the present paper we  study in detail this problem for the
spin-1/2 fermion perturbative self-energy and  the elastic EMV in
the Yukawa model within the framework of explicitly covariant
LFD~\cite{cdkm}. The latter deals with the LF plane of general
orientation $\omega\cd x= \omega_0 t-{\bg \omega}\cd {\bg x}=0$,
where $\omega$ is a four-vector with $\omega^2=0$. In the
particular case $\omega=(1,0,0,-1)$ we recover the standard
approach on the plane $t+z=0$. Due to $\omega$, which is
transformed as a four-vector under rotations and Lorentz boosts,
we can keep manifest rotational invariance throughout the
calculations. Dependence of amplitudes on the LF plane orientation
turns now in their dependence on the four-vector $\omega$. The
latter participates in the construction of  the  spin structures
in which the regularized initial LF amplitude can be decomposed,
on equal footing with  the particle four-momenta. This generates
extra ($\omega$-dependent) spin structures with corresponding
scalar coefficients (e.g., electromagnetic form factors). The
number of the extra spin structures and their explicit forms are
determined by general physical principles (more precisely, by
 the  particle spins and  the  symmetries of
the interaction), i.e., they are universal for any model and do
not depend on particular features of dynamics. Whereas, the
dependence of  the  extra coefficients on particle four-momenta is
determined by  the  model. This allows to separate general
properties, related to LFD itself, from model-dependent effects.

We shall proceed in the following two ways, both for the
self-energy and  the  EMV. In the first way, we calculate these
quantities by the LFD graph technique rules, taking into account
all necessary diagrams. In the second way, we start from the
standard Feynman amplitudes and integrate them in terms of the LF
variables.  For both ways, we have to introduce cutoffs on the LF
variables, and this fact already implies  the  contribution of
extra ($\omega$-dependent) structures and their corresponding form
factors. The regularized self-energy and the EMV calculated by
means of the LFD graph technique rules do not coincide in general
with their counterparts obtained  from the Feynman amplitudes. For
the EMV case, the vertex found from the Feynman amplitude with the
cutoffs imposed on the LF variables, also differs from that
calculated in a standard way, by the Wick rotation with a
spherically symmetric cutoff or by the Pauli-Villars (PV)
regularization. All these differences disappear when we deal with
integrals which are finite from the very beginning (e.g., due to
the PV regularization). They disappear also in the renormalized
amplitudes, though renormalization procedures (counterterms, etc.)
are drastically different for different regularization schemes.

Within covariant LFD, the perturbative QED self-energy and the EMV
in the channel of the fermion-antifermion pair creation have been
studied earlier~\cite{dugne}. The main subject was to extract the
physical ($\omega$-independent) contributions from the
corresponding amplitudes and to renormalize this physical part
only. The present analysis is devoted to a more detailed treatment
of the self-energy and  the  EMV. We calculate both physical and
non-physical contributions in the two ways mentioned above and
investigate the influence of the regularization procedure on the
whole amplitudes and on their subsequent renormalization.  The
Yukawa model which we use reflects some features of QED but it is
simpler from the technical point of view.

The paper is organized as follows. In Sec.~\ref{SE} we briefly
describe the LFD graph technique rules and apply them to calculate
the fermion self-energy. We use two different regularization
procedures, the transverse and longitudinal LF cutoffs or PV
subtractions, either for the bosonic, or simultaneously for both
bosonic and fermionic propagators.  Then we calculate the fermion
self-energy, starting from the manifestly invariant Feynman
amplitude expressed through the LF variables and regularized in
the same way as the LFD one. We compare the results obtained in
both approaches and  analyze  how they are affected by the choice
of regularization. In Sec.~\ref{EMV} we repeat analogous steps for
the fermion EMV. For this purpose, we derive the LF interaction
Hamiltonian which includes fermion-boson and fermion-photon
interactions.  We then  construct the complete  set of  the  LF
diagrams which contribute to the EMV. We use again the
non-invariant LF cutoffs and  the invariant PV regularization. The
LFD form factors are compared to those obtained in terms of the
Feynman amplitude with the same type of  regularization. General
discussion of our results is presented in Sec.~\ref{discuss}.
Sec.~\ref{concl} contains concluding remarks. The technical
details of some derivations are given in the Appendices.
%
\section{\label{SE} The fermion self-energy}
The fermion self-energy is the simplest example of how an extra
spin structure is generated by  rotationally non-invariant cutoffs
in LFD. To make the situation more transparent, we will calculate
the self-energy independently in the  two following ways: (1) by
applying the covariant LFD graph technique rules; (2) by using
the four-dimensional Feynman approach. In each case we consider
two different types of  regularization of divergent integrals:
either the traditional rotationally non-invariant cutoffs or  the
invariant PV regularization. We then renormalize the amplitudes
and compare the results obtained within these two methods.
\subsection{Calculation in light-front dynamics}
\subsubsection{\label{sec:LFDSE} Light-front diagrams and their amplitudes}
We calculate in  this section the fermion self-energy in the
second order of perturbation theory, using the graph technique
rules of explicitly covariant LFD~\cite{cdkm,kms_04,karm76}. We do
it in details in order to explain the rules on a concrete example.
\begin{figure}[btph]
\begin{center}
\includegraphics[scale=0.8]{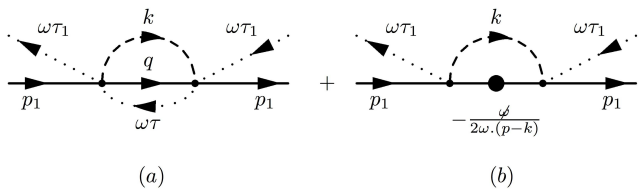}
\caption{\label{fig0a} Two contributions to the LFD fermion
self-energy $-\Sigma(p)$:  the  two-body intermediate state (a)
and the contact term (b). The solid, dashed, and dotted lines
represent, respectively, the fermion, the boson, and the spurion.
Here $p=p_1-\omega\tau_1$, where $\omega\tau_1$ is the
four-momentum attached to the initial (or final) spurion line. See
text for the explanation.}
\end{center}
\end{figure}

The self-energy $\Sigma(p)$ is determined by the sum of the two
diagrams shown in Fig.~\ref{fig0a},
\begin{equation}
\label{selfenLFD} \Sigma(p)=\Sigma_{2b}(p)+\Sigma_{fc}(p).
\end{equation}
They correspond to the two-body contribution and the fermion
contact term, respectively.  Analytical expressions for the
corresponding amplitudes  read
\begin{subequations}
\begin{eqnarray}\label{2b}
\Sigma_{2b}(p)&=&-\frac{g^2}{(2\pi)^3}\int \theta(\omega\cd
k)\delta(k^2-\mu^2)d^4k \nonumber\\
&&\phantom{-\frac{g^2}{(2\pi)^3}}
\times(\sla{q}+m)\theta(\omega\cd
q)\delta(q^2-m^2)d^4q \nonumber\\
&&\phantom{-\frac{g^2}{(2\pi)^3}}
\times\delta^{(4)}(p+\omega\tau-k-q) \frac{d\tau}{\tau-i0},
\\
\Sigma_{fc}(p)&=&\frac{g^2}{(2\pi)^3}\int
\frac{\sla{\omega}}{2\omega\cd(p-k)} \theta(\omega\cd
k)\delta(k^2-\mu^2)d^4k, \nonumber\\
&&\label{fc}
\end{eqnarray}
\end{subequations}
where $g$ is the coupling constant of the fermion-boson
interaction, $m$ and $\mu$ are the fermion and boson masses,
respectively.  In  the  covariant LFD graph technique, all the
four-momenta are on the corresponding mass shells. This is due to
the fact that the propagators are proportional to delta-functions:
$\theta(\omega\cd k)\delta(k^2-\mu^2)$ is the boson propagator,
$({\sla q}+m)\theta(\omega\cd q)\delta(q^2-m^2)$ is the fermion
one. Each theta-function $\theta(\omega\cd l)$ selects only one
value of $l_0=\sqrt{{\bf l}^2+m_l^2}$, of the two possible ones
allowed by the corresponding delta-function $\delta(l^2-m_l^2$).
There is no any conservation law for  the components of particle
four-momenta in the direction of $\omega$ (or for  the
minus-components, in the standard version of LFD). The
conservation is restored by the spurion  four-momentum
$\omega\tau$ which enters the delta-function
$\delta^{(4)}(p+\omega\tau-k-q)$. The factor $1/(\tau-i0)$ is the
spurion propagator and the $\tau$-integration is performed in
infinite limits. To avoid misunderstanding, we emphasize that
spurions are not true particles and do not affect particle
counting. They serve as a convenient way to describe the departure
of intermediate particles off the energy shell. The term "spurion"
itself is used for  shorter wording only. For this reason, the
intermediate state in the self-energy~(\ref{2b}) contains  one
fermion and one  scalar boson only. The self-energy is supposed to
be off-energy-shell, i.e., $\tau_1\neq 0$.

Integrating by means of the delta-functions over $d^4q$,
$d\tau$, and $dk_0$, we get
\begin{subequations}
\label{2b1fc1}
\begin{eqnarray}\label{2b1}
\Sigma_{2b}(p)&=&-\frac{g^2}{(2\pi)^3}\int
\frac{(\sla{p}-\sla{k}+\sla{\omega}\tau+m) \theta[\omega\cd(p-k)]}
{2\omega\cd(p-k)\tau}\frac{d^3k}{2\varepsilon_{\bf k}},
\nonumber\\
&&\\
\Sigma_{fc}(p)&=&\frac{g^2\sla{\omega}}{(2\pi)^3}\int
\frac{1}{2\omega\cd(p-k)} \frac{d^3k}{2\varepsilon_{\bf k}},
\label{fc1}
\end{eqnarray}
\end{subequations}
where $\varepsilon_{\bf k}\equiv k_0=\sqrt{{\bf k}^2+\mu^2}$ and
\begin{equation}
\label{tau_se} \tau=\frac{m^2-(p-k)^2}{2\omega\cd(p-k)}.
\end{equation}

Let us go over to the LF variables. First, we denote $x=(\omega\cd
k)/(\omega\cd p)$ (equivalent to $k^+/p^+$ in standard
non-covariant LFD on the surface $t+z=0$). We then split the
three-vector ${\bf k}$ into two parts: ${\bf k}={\bf
k}_{\perp}+{\bf k}_{\parallel}$, which are, respectively,
perpendicular and parallel to the three-vector ${\bg \omega}$.
Since $\Sigma(p)$ is an analytic function of $p^2$, we may
calculate it for $p^2>0$, while its values for $p^2\leq 0$ are
obtained by the analytical continuation. If $p^2>0$, we may
perform our calculation  in  the reference frame  where  ${\bf
p}={\bf 0}$. Using the kinematical relations from
Appendix~\ref{kin}, one can rewrite Eqs.~(\ref{2b1fc1}) as
\begin{widetext}
\begin{subequations}
\label{2b2fc2}
\begin{eqnarray}\label{2b2}
\Sigma_{2b}(p)&=&-\frac{g^2}{16\pi^3}\int d^2k_{\perp}\int_0^1
\frac{(\sla{p}-\sla{k}+m)dx}
{{\bf k}_{\perp}^2+m^2x-p^2x(1-x)+\mu^2(1-x)}
  -\frac{g^2{\sla \omega}}{32\pi^3(\omega\cd p)}\int
d^2k_{\perp}\int_0^{1}\frac{dx}{x(1-x)},\\
\Sigma_{fc}(p)&=&\frac{g^2{\sla \omega}}{32\pi^3(\omega\cd
p)}\int d^2k_{\perp}\int_0^{+\infty}\frac{dx}{x(1-x)}.
\label{fc2}
\end{eqnarray}
\end{subequations}
\end{widetext}
Both $\Sigma_{2b}(p)$ and $\Sigma_{fc}(p)$ are expressed through
integrals which diverge logarithmically in $x$ and quadratically
in $|{\bf k}_{\perp}|$. Possible regularization procedures are
discussed below.

Following Ref.~\cite{kms_04}, we will use the matrix
representation
\begin{subequations}
\label{sigdec}
\begin{eqnarray}
\Sigma_{2b}(p)&=&g^2\left[{\cal A}(p^2)+{\cal B}(p^2)\frac{{\sla
p}}{m}+{\cal C}(p^2)\frac{m{\sla \omega}}{\omega\cd p}\right],
\label{Sigdecomp2b} \\
\Sigma_{fc}(p)&=&g^2C_{fc}\frac{m{\sla \omega}}{\omega\cd p},
\label{Sigdecompfc}
\end{eqnarray}
\end{subequations}
where the coefficients ${\cal A}$, ${\cal B}$, and ${\cal C}$ are
scalar functions which depend on $p^2$ only. They are independent
of $\omega$. The coefficient $C_{fc}$ is a constant. The
self-energy is thus obtained by summing up Eqs.~(\ref{sigdec}):
\begin{equation}\label{Sigdecomp}
\Sigma(p)=g^2\left\{{\cal A}(p^2)+{\cal B}(p^2)\frac{{\sla
p}}{m}+[{\cal C}(p^2)+C_{fc}]\frac{m{\sla \omega}}{\omega\cd
p}\right\}.
\end{equation}
Note that in the expression~(\ref{Sigdecomp}) an additional spin
structure proportional to ${\sla \omega}$ appears,  as compared to
the standard four-dimensional Feynman approach.
\subsubsection{\label{regLd} Regularization with rotationally
non-invariant cutoffs}
In order to regularize the integrals  over $d^2k_{\perp}$ , we
introduce a cutoff $\Lambda_{\perp}$, so that ${\bf
k}_{\perp}^2<\Lambda_{\perp}^2$. Since some integrals over $dx$
diverge logarithmically at $x=0$ and/or $x=1$, we also introduce
(where it is needed) an infinitesimal positive cutoff $\epsilon$,
assuming that $x$ may belong to the intervals
$\epsilon<x<1-\epsilon$ and $1+\epsilon<x<+\infty$. The
corresponding analytical expressions for the functions ${\cal
A}(p^2)$, ${\cal B}(p^2)$, ${\cal C}(p^2)$, and $C_{fc}$ were
found in Ref~\cite{kms_04}:
\begin{widetext}
\begin{subequations}
\label{ABC2}
\begin{eqnarray}
\label{Ap2} {\cal
A}(p^2)&=&-\frac{m}{16\pi^2}\int_0^{\Lambda_{\perp}^2}d{\bf k}_{\perp}^2\int_0^1
dx\,\frac{1}{{\bf k}_{\perp}^2+m^2x-p^2x(1-x)+\mu^2(1-x)},
\\
\label{Bp2} {\cal
B}(p^2)&=&-\frac{m}{16\pi^2}\int_0^{\Lambda_{\perp}^2}d{\bf k}_{\perp}^2\int_0^1
dx\,\frac{1-x}{{\bf k}_{\perp}^2+m^2x-p^2x(1-x)+\mu^2(1-x)},
\\
\label{Cp2} {\cal
C}(p^2)&=&-\frac{1}{32\pi^2m}\int_0^{\Lambda_{\perp}^2}d{\bf k}_{\perp}^2\int_0^{1-\epsilon}
dx\,\frac{k_{\perp}^2+m^2-p^2(1-x)^2}{(1-x)[{\bf k}_{\perp}^2+m^2x-p^2x(1-x)+\mu^2(1-x)]},
\\
\label{Cfc}
C_{fc}&=&\frac{1}{32\pi^2m}\int_0^{\Lambda_{\perp}^2}d{\bf k}_{\perp}^2
\left[\int_{\epsilon}^{1-\epsilon}\frac{dx}{x(1-x)}+
\int_{1+\epsilon}^{+\infty}\frac{dx}{x(1-x)}\right].
\end{eqnarray}
\end{subequations}
\end{widetext}
We imply in the following that
$\Lambda_{\perp}^2\gg \mbox{max}\{|p^2|,m^2,\mu^2\}$ and
$\epsilon\ll 1$.  The dependence of physical results  on
the cutoffs is eliminated by taking the limits
\mbox{$\Lambda_{\perp}\to\infty$, $\epsilon\to 0$}. Retaining in
Eqs.~(\ref{ABC2}) all terms which do not vanish in
these limits, we get
\begin{widetext}
\begin{subequations}
\label{ABCp2}
\begin{eqnarray}
\label{Ap2a} {\cal
A}(p^2)&=&-\frac{m}{8\pi^2}\log\frac{\Lambda_{\perp}}{m}+\frac{m}{16\pi^2}
\int_0^1dx\,\log\left[\frac{m^2x-p^2x(1-x)+\mu^2(1-x)}{m^2}\right],
\\
\label{Bp2a} {\cal
B}(p^2)&=&-\frac{m}{16\pi^2}\log\frac{\Lambda_{\perp}}{m}+\frac{m}{16\pi^2}
\int_0^1dx\,(1-x)\log\left[\frac{m^2x-p^2x(1-x)+\mu^2(1-x)}{m^2}\right],
\\
\label{Cp2a} {\cal
C}(p^2)&=&-\frac{\Lambda_{\perp}^2}{32\pi^2m}\log\frac{1}{\epsilon}-
\frac{m^2-\mu^2}{16\pi^2m}\log\frac{\Lambda_{\perp}}{m}-\frac{1}{32\pi^2m}
\left(m^2-\mu^2-2\mu^2\log\frac{m}{\mu}\right),
\\
\label{Cfca}
C_{fc}&=&\phantom{-}\frac{\Lambda_{\perp}^2}{32\pi^2m}\log\frac{1}{\epsilon}.
\end{eqnarray}
\end{subequations}
\end{widetext}
We have not integrated over $dx$ in Eqs.~(\ref{Ap2a})
and~(\ref{Bp2a}) because the results of  the  integrations
are rather long. It is interesting to note that ${\cal C}(p^2)$
does not depend on $p^2$. Due to this, we will denote in the following ${\cal
C}(p^2)\equiv C=const$ and, for shortness, $\tilde{C}\equiv
C+C_{fc}$.

Each of the two quantities, $C$ and $C_{fc}$ diverges  like
$\Lambda_{\perp}^2\log(1/\epsilon)$.  The strongest divergencies
cancel in the sum $\tilde{C}=C+C_{fc}$, but the latter differs
from zero and, moreover, has no finite limit when
$\Lambda_{\perp}\to\infty$:
\begin{equation}
\label{Cp2b} \tilde{C}=-
\frac{m^2-\mu^2}{16\pi^2m}\log\frac{\Lambda_{\perp}}{m}-\frac{1}{32\pi^2m}
\left(m^2-\mu^2-2\mu^2\log\frac{m}{\mu}\right).
\end{equation}
Since the two diagrams shown in Fig.~\ref{fig0a} exhaust the full
set of the second-order diagrams which contribute to the fermion
self-energy, we might expect  the  disappearance of every
${\omega}$- dependent contribution from the amplitude. As far as
this does not take place, the only source of the
$\omega$-dependence is the use of  the  rotationally non-invariant
cutoffs for the LF variables ${\bf k}_{\perp}^2$ and $x$. Indeed,
if we write  these variables  in  the explicitly covariant form
$$
{\bf k}_{\perp}^2=2\frac{(\omega\cd k)(k\cd p)}{\omega\cd p}-
p^2\left(\frac{\omega\cd k}{\omega\cd
p}\right)^2-\mu^2,\,\,\,\,\,\,\,\,x=\frac{\omega\cd k}{\omega\cd
p},
$$
it becomes evident that both of them depend on $\omega$.
Introducing the cutoffs $\Lambda_{\perp}^2$ and $\epsilon$, we
restrict an $\omega$-dependent integration domain, which
inevitably brings $\omega$-dependence into the regularized
quantities. We will demonstrate this feature in more detail in
Sec.~\ref{discuss} by using a very simple and transparent example.
Note that without adding  new, $\omega$-dependent, counterterms
in the interaction Hamiltonian,  this dependence is not killed by
the standard renormalization. The renormalization recipe must be
therefore modified~\cite{kms_04}.
\subsubsection{Invariant Pauli-Villars regularization}
As we learned above, the source of  the appearance of the extra
($\omega$-dependent) term in the regularized fermion self-energy
is the $\omega$-dependence of the integration domain. A standard
way free from this demerit is the use of the PV regularization,
since in that case the cutoffs have no more relation to $\omega$.
In the language of LFD, the PV regularization  consists in
changing the propagators as
\begin{equation*}
\theta(\omega\cd k)\delta(k^2-\mu^2)\to\theta(\omega\cd
k)[\delta(k^2-\mu^2)-\delta(k^2-\mu_1^2)]
\end{equation*}
for scalar bosons, and
\begin{multline*}
({\sla q}+m)\theta(\omega\cd q)\delta(q^2-m^2)\to\\
\theta(\omega\cd
q)\left[({\sla q}+m)\delta(q^2-m^2)
 -({\sla q}+m_1)\delta(q^2-m_1^2)\right]
\end{multline*}
for fermions.  This procedure is equivalent to introducing
additional particles (one PV fermion with the mass $m_1$ and one
PV boson with the mass $\mu_1$), whose  wave functions have
negative norms.  If  needed,  more subtractions can be done till
all integrals become convergent. After the calculation of the
integrals and the renormalization, the limits $\mu_1\to\infty$ and
$m_1\to\infty$ should be taken. In the case of the fermion
self-energy, we may regularize either  the boson propagator only,
or  the  fermion one, or both simultaneously. Hereafter we will
supply PV-regularized quantities with the superscript "$PV,\,b$"
(when only  the  bosonic propagator is modified) or "$PV,\,b+f$"
(when both bosonic and fermionic propagators are modified).

Let us now calculate the PV-regularized coefficients ${\cal
A}(p^2)$, ${\cal B}(p^2)$, and $\tilde{C}$. We  can start from the
expressions~(\ref{ABCp2}), in spite of their dependence on the
"old" cutoffs $\Lambda_{\perp}$ and $\epsilon$. Indeed, the
integrals in Eqs.~(\ref{2b2fc2}) become regular, provided enough
PV subtractions have been made. If so, they have definite limits
at $\Lambda_{\perp}\to \infty$ and $\epsilon\to 0$.

The integrals for ${\cal A}(p^2)$ and ${\cal B}(p^2)$ become
convergent after the regularization by a PV boson only:
\begin{widetext}
\begin{subequations}
\label{ABPV}
\begin{eqnarray}
\label{APV} {\cal A}^{PV,\,b}(p^2) &=& {\cal A}(p^2,m,\mu)-{\cal
A}(p^2,m,\mu_1)=\frac{m}{16\pi^2}
\int_0^1dx\,\log\left[\frac{m^2x-p^2x(1-x)+\mu^2(1-x)}
{m^2x-p^2x(1-x)+\mu_1^2(1-x)}\right],
\\
\label{BPV}
 {\cal B}^{PV,\,b}(p^2)&=&{\cal B}(p^2,m,\mu)-{\cal B}(p^2,m,\mu_1)
=\frac{m}{16\pi^2}
\int_0^1dx\,(1-x)\log\left[\frac{m^2x-p^2x(1-x)+\mu^2(1-x)}
{m^2x-p^2x(1-x)+\mu_1^2(1-x)}\right].
\end{eqnarray}
\end{subequations}
\end{widetext}

The situation differs drastically for the coefficient $\tilde{C}$.
After  the bosonic  PV regularization we get
\begin{multline}
\label{Cp2CfcPVb}
\tilde{C}^{PV,\,b}=\frac{1}{32\pi^2m}\left[(\mu^2-\mu_1^2)\left(1+2\log
\frac{\Lambda_{\perp}}{m}\right)\right.\\
+\left. 2\mu^2\log\frac{m}{\mu}-2\mu_1^2\log\frac{m}{\mu_1}\right].
\end{multline}
Since the result is still divergent for
$\Lambda_{\perp}\to\infty$, this regularization is not enough. The
additional fermionic  PV regularization requires some care because
$\tilde{C}$ is a coefficient at the spin structure $m{\sla
\omega}/(\omega\cd p)$ which itself depends on $m$. Hence, one
should regularize the quantity $m\tilde{C}$:
\begin{multline}
\label{Cp2CfcPVbf} \left(m\tilde{C}\right)^{PV,\,b+f}=
m\tilde{C}(m,\mu)-m\tilde{C}(m,\mu_1)\\
-m_1\tilde{C}(m_1,\mu)+m_1\tilde{C}(m_1,\mu_1)=0.
\end{multline}
We see that after the double PV regularization the extra structure
in Eq.~(\ref{Sigdecomp}), proportional to ${\sla \omega}$,
disappears, as it should. Note that Eq.~(\ref{Cp2CfcPVbf}) holds
for arbitrary (i.e., not necessary infinite) PV masses $m_1$ and
$\mu_1$.
\subsubsection{Renormalization procedure}
The renormalized  self-energy is obtained by using the
standard procedure:
\begin{equation}
\label{sigmaren} \Sigma_{ren}^{PV,\,b+f}(p)=\Sigma^{PV,\,b+f}(p)-c_1-c_2({\sla p}-m),
\end{equation}
where
\begin{subequations}
\begin{eqnarray}
\label{c12}
c_1&=&\left.\frac{\bar{u}(p)\Sigma^{PV,\,b+f}(p)u(p)}{2m}\right|_{p^2=m^2},\\
c_2&=&\frac{1}{2m}\left\{\bar{u}(p)\frac{\partial
\Sigma^{PV,\,b+f}(p)}{\partial {\sla p}}u(p)\right\}_{p^2=m^2}.
\end{eqnarray}
\end{subequations}
Finally,
\begin{equation}
\label{sigmafinren} \Sigma_{ren}^{PV,\,b+f}(p)=g^2\left[{\cal
A}_{ren}(p^2)+{\cal B}_{ren}(p^2)\frac{\sla p}{m}\right],
\end{equation}
with
\begin{subequations}
\begin{eqnarray}
{\cal A}_{ren}(p^2)={\cal A}(p^2)-{\cal
A}(m^2)+2m^2[{\cal A}'(m^2)+{\cal B}'(m^2)],\nonumber \\
\label{Aren} \\
{\cal B}_{ren}(p^2)={\cal B}(p^2)-{\cal
B}(m^2)-2m^2[{\cal A}'(m^2)+{\cal B}'(m^2)].\nonumber\\
\label{Bren} \end{eqnarray}
\end{subequations}
In order to calculate ${\cal A}_{ren}(p^2)$ and ${\cal
B}_{ren}(p^2)$ we can  use the initial functions ${\cal A}(p^2)$
and ${\cal B}(p^2)$ regularized either by the non-invariant
cutoffs, Eqs.~(\ref{Ap2a}) and~(\ref{Bp2a}), or by means of the PV
regularization, Eqs.~(\ref{ABPV}).  Any choice leads  to the same
result:
\begin{subequations}
\label{ABren}
\begin{eqnarray}
\label{Aren1} {\cal A}_{ren}(p^2)&=&\frac{m}{16\pi^2}\int_0^1 dx
\left[\phi_1(x)-\phi_2(x)\right],
\\
{\cal B}_{ren}(p^2)&=&\frac{m}{16\pi^2}\int_0^1
dx\,(1-x) \left[\phi_1(x)+\phi_2(x)\right],\nonumber \\
\label{Bren1}
\end{eqnarray}
\end{subequations}
where
\begin{eqnarray*}
\phi_1(x)&=&\log\left[\frac{m^2x-p^2x(1-x)+\mu^2(1-x)}{m^2x^2+\mu^2(1-x)}\right],
\\
\phi_2(x)&=&\frac{2m^2x(2-3x+x^2)}{m^2x^2+\mu^2(1-x)}.
\end{eqnarray*}
The remaining integrations over $dx$ in Eqs.~(\ref{ABren})
are simple but lengthy.
\subsection{\label{SEF} Calculation in  the  four-dimensional \\ Feynman
approach}
\subsubsection{\label{Sigma_Fey} Regularization with rotationally
non-invariant cutoffs} We showed above that the regularized
fermion self-energy calculated within LFD with the traditional
transverse ($\Lambda_{\perp}$) and longitudinal ($\epsilon$)
cutoffs contains an extra spin structure depending on $\omega$.
In order to understand the reasons of this behavior,  we
calculate here the self-energy in another way, following
Ref.~\cite{lb95}. We start from the standard four-dimensional
Feynman expression
\begin{equation}
\label{sef} \Sigma_F(p)=\frac{ig^2}{(2\pi)^4}\int d^4 k\;
\frac{{\sla p}-{\sla
k}+m}{[k^2-\mu^2+i0][(p-k)^2-m^2+i0]},
\end{equation}
but perform the integrations in terms of the LF variables, with
the corresponding cutoffs. For this purpose, we introduce the
minus-, plus-, and transverse components of the four-momentum $k$:
$$
k_-=k_0-k_z,\quad k_+=k_0+k_z,\quad {\bf k}_{\perp}=(k_x,k_y),
$$
and analogously for $p$. As in Sec.~\ref{sec:LFDSE}, we take, for
convenience, the reference frame where ${\bf p}={\bf 0}$. In this
frame $p_-=p^2/p_+$. Denoting $k_+=x p_+$, we get
\begin{multline}
\label{sigma} \Sigma_F(p)= \frac{ig^2p_+}{32\pi^4}\int d^2
k_{\perp}\int_{-\infty}^{+\infty}dx \\
\times\int_{-\infty}^{+\infty}dk_-
\frac{1}{[k_- p_+ x -{\bf k}_{\perp}^2-\mu^2+i0]}
\\
\times\frac{({\sla p}-{\sla k}+m)} {[(p_- -k_-)p_+(1-x)-{\bf k}_{\perp}^2
 -m^2+i0]}
\end{multline}
with
$$
{\sla k}=\frac{1}{2}\gamma_+k_-+\frac{x}{2}\gamma_-p_+-{\bg
\gamma}_{\perp}\cd {\bf k}_{\perp}.
$$
The integral over $dk_-$ is calculated by using the principal
value prescription:
\begin{equation}
\label{Lkm}
\int_{-\infty}^{+\infty}dk_-(\ldots)=\lim_{L\to\infty}\int_{-L}^Ldk_-(\ldots).
\end{equation}
This integral is well defined unless $x=0$ or $x=1$. If $x=0$ or
$x=1$, the integral (\ref{Lkm}) diverges on the upper and lower
limits. So, the infinitesimal integration domains  near the points
$x=0$ and $x=1$ (the  so-called zero modes) require special
consideration. We introduce a cutoff $\epsilon$ in the variable
$x$ and represent $\Sigma(p)$ as a sum
\begin{equation}
\label{sigmap} \Sigma_F(p)=\Sigma_{p+a}(p)+\Sigma_{zm}(p),
\end{equation}
where $\Sigma_{p+a}(p)$ incorporates  the  contributions from the
regions of integration over $dx$ ($-\infty < x < \infty$),
excluding the singular points $x=0$ and $x=1$, while the zero-mode
part $\Sigma_{zm}(p)$ involves the integrations in the
$\epsilon$-vicinities of these two points.

The calculation is carried out in Appendix~\ref{sen}. After
closing the integration contour by an arc of a circle (see
Appendix~\ref{sen}), the integral for $\Sigma_{p+a}(p)$ is
represented as a sum of the pole and arc contributions and has the
form
\begin{multline}
\label{SigmaA1} \Sigma_{p+a}(p)=-\frac{g^2}{16\pi^3}\int
d^2k_{\perp}\\
\times\int_{\epsilon}^{1-\epsilon} \frac{({\sla p}-{\sla
k}+m)dx}{{\bf k}_{\perp}^2+m^2x-p^2x(1-x)+\mu^2(1-x)}
\\
 +\frac{g^2{\sla \omega}}{32\pi^3(\omega\cd p)}\int
d^2k_{\perp}\int_{1+\epsilon}^{+\infty}\frac{dx}{x(1-x)}.
\end{multline}
Comparing Eq.~(\ref{SigmaA1}) with Eqs.~(\ref{2b2fc2}), we see
that
\begin{equation}
\label{SigmaA2} \Sigma_{p+a}(p)=\Sigma_{2b}(p)+\Sigma_{fc}(p),
\end{equation}
where both terms on the right-hand side are regularized at $x=0$
and $x=1$. The pole plus arc contributions to Eq.~(\ref{sigma})
reproduce the result given by the sum of the two LFD diagrams
shown in Fig.~\ref{fig0a}. Hence,
\begin{equation}
\label{SigmaA3} \Sigma_{p+a}(p)=g^2\left[{\cal A}(p^2)+{\cal
B}(p^2)\frac{\sla p}{m}+C_{p+a}\frac{m{\sla \omega}}{\omega\cd
p}\right],
\end{equation}
where ${\cal A}(p^2)$ and ${\cal B}(p^2)$ are defined by
Eqs.~(\ref{Ap2a}) and~(\ref{Bp2a}), while $C_{p+a}$ coincides with
$\tilde{C}$, Eq.~(\ref{Cp2b}).

As mentioned above, the zero mode contribution results from the
divergence (at $L\to\infty$) of the integral over $k_-$, which
occurs when $x=0,1$. This divergence is determined by the leading
$k_-$-term in the numerator of Eq.~(\ref{sigma}), that is by
$\frac{1}{2}\gamma_+k_-$. In explicitly covariant LFD, the matrix
$\gamma_+$ turns  into ${\sla \omega}$.
The zero-mode contribution is therefore
\begin{equation}
\label{sigma0}
 \Sigma_{zm}(p)\equiv g^2C_{zm}\frac{m{\sla \omega}}{\omega\cd p},
\end{equation}
where $C_{zm}$ is calculated in Appendix \ref{sen} and has the
form
\begin{eqnarray}
\label{CZM}
 C_{zm} &=&
\frac{m^2-\mu^2}{16\pi^2m}\log\frac{\Lambda_{\perp}}{m} \\
&&+\frac{1}{32\pi^2m}
\left(m^2-\mu^2-2\mu^2\log\frac{m}{\mu}\right).\nonumber
\end{eqnarray}
We thus find
\begin{equation}\label{CFeynman}
C_{p+a}+C_{zm}=0,
\end{equation}
since $C_{p+a}=\tilde{C}$ is given by the right-hand side of
Eq.~(\ref{Cp2b}). Substituting Eqs.~(\ref{SigmaA3})
and~(\ref{sigma0}) into Eq.~(\ref{sigmap}), and taking into
account Eq.~(\ref{CFeynman}), we finally get
\begin{equation}
\label{sigmafin} \Sigma_F(p)=g^2\left[{\cal A}(p^2)+{\cal
B}(p^2)\frac{{\sla p}}{m}\right].
\end{equation}

We see that after the incorporation of the pole, arc, and
zero-mode contributions, the $\omega$-dependent term in the
self-energy disappears, even before the renormalization. Note that
although we used the same cutoffs, the formula~(\ref{sigmafin})
does not coincide with the expression~(\ref{Sigdecomp}) obtained
by using the LFD graph technique rules, since the sum ${\cal
C}(p^2)+C_{fc}$ is not zero [it is given by Eq.~(\ref{Cp2b})]. One
might think that the dependence of the
self-energy~(\ref{Sigdecomp}) on $\omega$ is an artefact of the
LFD rules, whereas the independence of the Feynman approach on
$\omega$ is natural, since we started with the Feynman
expression~(\ref{sef}) which "knows nothing" about $\omega$. It is
not so, since the initially divergent integral for $\Sigma(p)$
acquires some sense only after regularization, and the latter has
been done in terms of the cutoffs imposed on the LF variables. The
independence of $\Sigma_F(p)$ on $\omega$ looks thereby as a
coincidence. We shall see in Sec.~\ref{EMFey} that, in the EMV
case, in contrast to the self-energy  one, the LF cutoffs applied
to the initial Feynman integral {\em do result in} some dependence
of the EMV on the LF plane orientation.

After the renormalization, Eq.~(\ref{sigmafin})  reproduces the
expression~(\ref{sigmafinren}) obtained earlier for the
self-energy found within LFD  and regularized by the invariant PV
method.
\subsubsection{\label{ir} Invariant regularization}
We briefly recall in this section familiar results known from the
standard four-dimensional Feynman formalism which is completely
independent from LFD. It may serve as an additional test of the
results obtained above.

The fermion self-energy is given by Eq.~(\ref{sef}). It is
convenient to use the following decomposition:
\begin{equation}\label{sefF}
\Sigma_F(p)=g^2\left[{\cal A}_F(p^2)+{\cal B}_F(p^2)\frac{\sla
p}{m}\right],
\end{equation}
where ${\cal A}_F(p^2)$ and ${\cal B}_F(p^2)$ are scalar
functions.  Note that they do not coincide with ${\cal
A}(p^2)$ and ${\cal B}(p^2)$ from Eq.~(\ref{sigmafin}), since we
use here another regularization procedure.  Applying the
Feynman parametrization, we can rewrite Eq.~(\ref{sef}) as
\begin{multline}\label{sef1}
\Sigma_F(p)=\frac{ig^2}{16\pi^4}\int d^4k'\\
\times\int_0^1 dx\,
\frac{(1-x){\sla p}-{\sla
k}'+m}{[{k'}^2-m^2x+p^2x(1-x)-\mu^2(1-x)+i0]^2},
\end{multline}
where we introduced $k'=k-xp$. Going over to Euclidean space by
means of the Wick rotation  $k'_0=ik'_4$ with real $k_4'$ and
regularizing the divergent integrals by an invariant cutoff
$|{k'}^2|={{\bf k}'}^2+{k_4'}^2<\Lambda^2$ (assuming that
$\Lambda^2\gg \{m^2,\mu^2,|p^2|\}$), we can perform the
four-dimensional integration and get
\begin{multline}\label{sef2}
\Sigma_F(p)=-\frac{g^2}{16\pi^2}\int_0^1 dx\,[(1-x){\sla
p}+m]\\
\times\left\{
\log\left[\frac{\Lambda^2}{m^2x-p^2x(1-x)+\mu^2(1-x)}\right]-1\right\}.
\end{multline}
The terms of order $1/\Lambda^2$ and higher are omitted. Comparing
the right-hand sides of Eqs.~(\ref{sefF}) and~(\ref{sef2}), we
find
\begin{widetext}
\begin{subequations}
\begin{eqnarray}
\label{Ap2F} {\cal
A}_F(p^2)&=&-\frac{m}{16\pi^2}\left(\log\frac{\Lambda^2}{m^2}-1\right)
+
\frac{m}{16\pi^2}\int_0^1 dx\,
\log\left[\frac{m^2x-p^2x(1-x)+\mu^2(1-x)}{m^2}\right],
\\
\label{Bp2F} {\cal
B}_F(p^2)&=&-\frac{m}{32\pi^2}\left(\log\frac{\Lambda^2}{m^2}-1\right)
+\frac{m}{16\pi^2}\int_0^1 dx\,
(1-x)\log\left[\frac{m^2x-p^2x(1-x)+\mu^2(1-x)}{m^2}\right].
\end{eqnarray}
\end{subequations}
\end{widetext}

The renormalized self-energy $\Sigma_{F,\,ren}(p)$ is found from
Eq.~(\ref{sigmaren}), changing $\Sigma^{PV,\,b+f}(p)$ by
$\Sigma_{F}(p)$. The corresponding renormalized functions ${\cal
A}_{F,\,ren}(p^2)$  and ${\cal B}_{F,\,ren}(p^2)$ coincide with
${\cal A}_{ren}(p^2)$ and ${\cal B}_{ren}(p^2)$ given by
Eqs.~(\ref{ABren}). We  thus reproduce again
Eq.~(\ref{sigmafinren})  for $\Sigma_{F,\,ren}(p)$ .

It is easy to verify that using the PV subtraction for the boson
propagator (instead of the cutoff $\Lambda^2$),
$$
\Sigma_F^{PV,\,b}(p)=\Sigma_F(p,m,\mu)-\Sigma_F(p,m,\mu_1),
$$
leads, in the limit $\mu_1\to\infty$, to the same
expression~(\ref{sigmafinren}) for the renormalized self-energy.
We have therefore
\begin{equation}
\label{sigmaFLFD}
\Sigma_{F,\,ren}(p)=\Sigma_{F,\,ren}^{PV,\,b}(p)=\Sigma_{ren}^{PV,b+f}(p).
\end{equation}

To conclude, any of the considered ways of regularization (either
with the non-invariant cutoffs $\Lambda_{\perp}$ and $\epsilon$ or
with the single bosonic PV subtraction) can be used in order to
get the correct expression for the renormalized self-energy in the
Feynman approach, while the double (bosonic + fermionic) PV
subtraction is required in the  case of LFD.
\section{\label{EMV} Fermion electromagnetic vertex}
We can now proceed to the calculation of the spin-1/2 fermion
elastic EMV, $\Gamma_{\rho}$, which is connected with the matrix
element of the electromagnetic current $J_{\rho}$ by the relation
\begin{equation}
\label{eq0} J_{\rho}=e\bar{u}(p')\Gamma_{\rho}u(p),
\end{equation}
where $e$ is the physical electromagnetic coupling constant, $p$
and $p'$ are the initial and final on-mass-shell fermion
four-momenta (${p'}^2=p^2=m^2$). Assuming   $P$-,
$C$-, and $T$-parity conservation, $\Gamma_{\rho}$ is defined by
two form factors depending on the momentum transfer squared
$Q^2=-(p'-p)^2$:
\begin{equation}
\label{FFs}
\bar{u}'\Gamma_{\rho}u=\bar{u}'\left[F_1(Q^2)\gamma_{\rho}+
\frac{iF_2(Q^2)}{2m}\sigma_{\rho\nu}q_{\nu}\right]u.
\end{equation}
We omit for simplicity the bispinor arguments and denoted $q=p'-p$,
$\sigma_{\rho\nu}=i(\gamma_{\rho}\gamma_{\nu}-\gamma_{\nu}\gamma_{\rho})/2$.

We consider here, as in Sec.~\ref{SE}, the Yukawa model which
takes into account interaction of fermions with scalar bosons,
while  the  EMV "dressing" due to fermion-photon interactions is
neglected.  The  fermion-boson interaction is treated
perturbatively, up to terms of order $g^2$.

At that order, the EMV must be renormalized. The standard
renormalization recipe consists in  the subtraction
$\Gamma_{\rho}^{ren}=\Gamma_{\rho}-Z_1\gamma_{\rho}$ with the
constant $Z_1$ found from the requirement
$\bar{u}(p)\Gamma_{\rho}^{ren}u(p)=\bar{u}(p)\gamma_{\rho}u(p)$.
This leads to the following well-known expressions for the
renormalized form factors:
\begin{equation}
\label{FFren}
F_1^{ren}(Q^2)=1+F_1(Q^2)-F_1(0),\,\,\,\,\,\,\,\,F_2^{ren}(Q^2)=F_2(Q^2).
\end{equation}

We shall  follow the same ideology that we exposed above for the
self-energy: independent calculations of the EMV are performed,
within covariant LFD and the Feynman approach, both for
non-invariant and invariant regularization. However, in contrast
to the self-energy case, the use of rotationally non-invariant
cutoffs results in  the appearance of new structures (and  form
factors) in the EMV, even if one starts from the standard
four-dimensional Feynman expression. By this reason, the
renormalization of the two physical form factors only, as
prescribed by Eqs. (\ref{FFren}), is not enough to get the full
renormalized EMV.
\subsection{\label{EMvLFD} Calculation in light-front dynamics}
\subsubsection{\label{ham} Light-front interaction Hamiltonian involving
electromagnetic interaction}

Before going over to the consideration of the EMV in covariant LFD, one
should derive the interaction Hamiltonian which involves both
fermion-boson and fermion-photon interactions. Although all the
diagrams calculated below are generated by the graph technique
rules formulated in Ref.~\cite{cdkm}, this is another way to
explain their origin.

We will not give here a detailed derivation. The corresponding
procedure is exposed in Ref.~\cite{kms_04}. We derived there the
LF Hamiltonian describing a system of interacting fermion and
massless vector boson fields. This expression holds also for a
fermion-photon system. When the photon field is taken in the
Feynman gauge,  the Hamiltonian in Schr\"{o}dinger representation
has the form
\begin{equation}
\label{ham1} H^{int}(x)=-e\bar{\psi}\slaA\psi
+e^2\bar{\psi}\slaA\frac{\sla{\omega}}{2i(\omega\cd
D_A)}\slaA\psi,
\end{equation}
where $\psi$ and $A$ are the free fermion and photon fields,
respectively, and the operator $1/(i\omega\cd D_A)$ in coordinate
space acts on the coordinate along the four-vector $\omega$ (for
convenience we denote it for a moment as $x_-$, since $\omega$ in
standard LFD has only  the  minus-component):
\begin{multline} \label{1/D} \frac{1}{i\omega\cd
D_A}f(x_-)= \exp\left[-\frac{e}{i\omega\cd\partial}\omega\cd
A\right]\\
\times\frac{1}{i\omega\cd\partial}
\left\{\exp\left[\frac{e}{i\omega\cd\partial}\omega\cd
A\right]f(x_-)\right\}
\end{multline}
and  $1/(i\omega\cd
\partial)$ is the free reversal derivative operator:
\begin{equation}
\label{1/d} \frac{1}{i\omega\cd \partial}f(x_-)=-\frac{i}{4} \int
dy_-\epsilon(x_--y_-)f(y_-),
\end{equation}
where $\epsilon(x)$ is the sign function.
The operators $1/(i\omega\cd\partial)$ inside the exponents act on
the functions $\omega\cd A$ only, while that standing between the
exponents acts on all the functions to the right of it. In
momentum representation, the action of the operator $1/(i\omega\cd
\partial)$ on a function $f(x)$ reduces to the multiplication of
its Fourier transform $f(k)$ by the factor $1/(\omega\cd k)$.

It is easy to modify the Hamiltonian~(\ref{ham1}) in order to
incorporate interactions between fermions and scalar
bosons. The equation of motion for the Heisenberg fermion field
operator $\Psi$,  in the absence of scalar bosons,
looks like $(i{\sla\partial}-m)\Psi=-e{\slaCA}\Psi$, where
${\cal A}$ is the Heisenberg photon field operator. If we
introduce a scalar boson field $\Phi$, the equation
of motion becomes $(i{\sla\partial}-m)\Psi=(-e{\slaCA}-g\Phi)\Psi$.
Since the latter equation of motion is obtained
from the previous one by the substitution ${\slaCA}\to
{\slaCA}+(g/e)\Phi$, it is enough to make  the same
 substitution in the Hamiltonian~(\ref{ham1}),
everywhere except in the operator $1/(i\omega\cd D_A)$
~\cite{kms_04}.  The Hamiltonian thus becomes
\begin{eqnarray}
\label{ham2} H^{int}(x)=&-&\bar{\psi} \left[g{\varphi}+e{\slaA}\right] \psi \nonumber\\
&+&\bar{\psi} \left[g{\varphi}+e{\slaA}\right]
\frac{\hat{\omega}}{2i\omega\cd D_A} \left[g{\varphi}+e{\slaA}\right] \psi.
\end{eqnarray}
We see that the LFD interaction Hamiltonian~(\ref{ham2}) involves
also, besides the  usual term $-\bar{\psi}
\left[g{\varphi}+e{\slaA}\right] \psi$ describing ordinary
fermion-boson and fermion-photon interactions, the so-called
contact terms which are nonlinear in  the coupling constants. Note
that if we expand the Hamiltonian~(\ref{ham2}) in powers of $e$,
this expansion contains an infinite number of terms. Such a
peculiarity is connected with the photon spin and with the gauge
we have chosen.

In this paper, we are not interested in studying electromagnetic
effects, but focus on the interaction between fermions and scalar
bosons. We therefore restrict the Hamiltonian to the first order
in the electromagnetic coupling constant $e$, neglecting the terms
of order $e^2$ and higher. The result is
\begin{equation}
\label{hamtot} H^{int}=H_{fb1}+H_{fb2}+H_{em1}+H_{em2}+H_{em3},
\end{equation}
where
\begin{subequations}
\begin{eqnarray}
\label{hamfb1}
H_{fb1} & = & -g\bar{\psi}\psi\varphi,\\
\label{hamfb2}
H_{fb2} & = & g^2\bar{\psi}\varphi\frac{\sla
\omega}{2i\omega\cd\partial}\varphi\psi, \\
\label{hamem1}
H_{em1} & = & -e\bar{\psi}{\sla A}\psi,\\
\label{hamem2}
H_{em2} & = & eg\bar{\psi}\left(\varphi\frac{\sla
\omega}{2i\omega\cd\partial}{\sla A}+{\sla A}\frac{\sla
\omega}{2i\omega\cd\partial}\varphi\right)\psi,\\
H_{em3} & = & \frac{1}{2}eg^2\bar{\psi}\varphi
\left\{\frac{\sla
\omega}{i\omega\cd\partial}\left[\frac{1}{i\omega\cd\partial}\omega\cd
A\right]\right.\nonumber \\
\label{hamem3}
&&\left.- \left[\frac{1}{i\omega\cd\partial}\omega\cd
A\right]\frac{\sla \omega}{i\omega\cd\partial}\right\}\varphi\psi.
\end{eqnarray}
\end{subequations}
The contact terms are $H_{fb2}$, $H_{em2}$, and $H_{em3}$. The
operators $1/(i\omega\cd \partial)$ inside the squared brackets
act on $\omega\cd A$ only.
\subsubsection{\label{LFvert}Light-front diagrams and their amplitudes}
\begin{figure}[btph]
\begin{center}
\centerline{\includegraphics[scale=0.8]{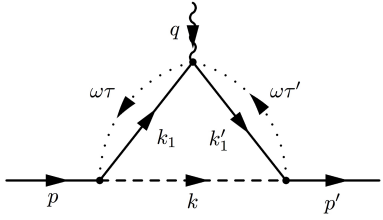}}
\caption{\label{triangle} Triangle LF diagram  }
\end{center}
\end{figure}
Since the amplitude of the process which we are interested in is
proportional to $eg^2$, we should collect together the matrix
elements from the Hamiltonian~(\ref{hamtot}) in the first, second
and third orders of perturbation theory. It can be
written schematically as
$$
<H^2_{fb1}H_{em1}>+<H_{fb1}H_{em2}>+<H_{em3}>.
$$
\begin{figure}[btph]
\begin{center}
\centerline{\includegraphics[scale=0.8]{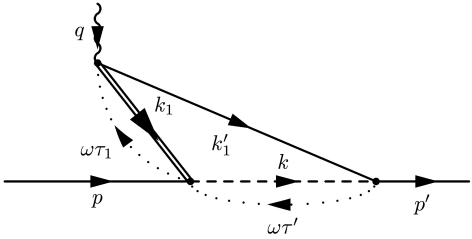}}
\caption{\label{pair} Pair creation by a photon. The double solid
line stands for an antifermion}
\end{center}
\end{figure}
Note that although the matrix element of the second order of
perturbation theory, $<H_{fb2}H_{em1}>$, is also of order $eg^2$,
it does not result in irreducible diagrams. The
Hamiltonian~(\ref{hamtot}) produces therefore the five
contributions to the EMV, shown in
Figs.~\ref{triangle}--\ref{double_ct}. The triangle and pair
creation diagrams are generated by ${<H^2_{fb1}H_{em1}>}$, the
left and right contact terms come from $<H_{fb1}H_{em2}>$, while
$<H_{em3}>$ is responsible for the double contact term.
\begin{figure}[btph]
\begin{center}
\centerline{\includegraphics[scale=0.7]{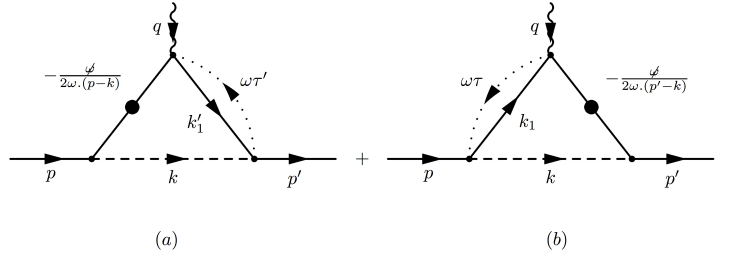}} \caption{\label{lrct}
Left (a) and right (b) contact terms}
\end{center}
\end{figure}
Applying the rules of the LFD graph technique~\cite{cdkm} to these
diagrams \footnote{The  rules incorrectly  prescribe to use  the
theta-function  $\theta(\omega\cd k)$  for the contact term. This
theta-function should be removed.}, we can find analytical
expressions for the corresponding amplitudes. However, before
writing  them down, one should note the following. When
calculating the form factors in covariant LFD, the condition
$\omega\cd q=0$ on the momentum transfer is usually imposed
(equivalent  to $q_+=q_0+q_z=0$ in  the non-covariant version of
LFD). It comes from the analysis of the pure scalar "EMV" (i.e.,
when all the particles, including the photon, are spinless), where
it forbids the pair creation diagram. Indeed, since the
plus-component of the pair momentum is always positive, the pair
can not be created by a virtual photon with $q_+=0$.  As a result,
when $q_+\to 0$, the corresponding phase space volume tends to
zero, and the amplitude of the pair creation diagram disappears.
For systems involving fermions and/or vector photons, the
amplitude of the pair creation diagram becomes indefinite if
$\omega\cd q$ exactly equals zero, since  it is given by an
integral with  an  infinitely large integrand and   zero phase
space volume. For this reason, one has to take $\omega\cd q\neq
0$.  We set $\omega\cd q\equiv \alpha(\omega\cd p)$, where
$\alpha$ is a constant which we take positive, for definiteness.
We will see below that for the rotationally non-invariant cutoffs
discussed in Sec.~\ref{SE}, a non-zero contribution to the EMV
from the pair creation diagram survives, and moreover, it tends to
infinity when $\alpha\to 0$.
\begin{figure}[btph]
\begin{center}
\centerline{\includegraphics[scale=0.8]{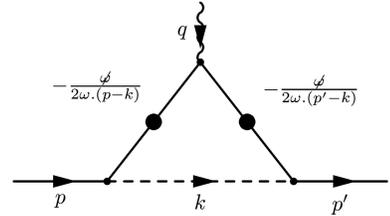}} \caption{\label{double_ct}
Double contact term}
\end{center}
\end{figure}

We can now proceed to  the calculation of  the  LF
diagram amplitudes.  The contribution of the triangle diagram on
Fig.~\ref{triangle}   reads
\begin{eqnarray}
\label{GammaA} \Gamma_{\rho}^{(tri)}&= &\frac{g^2}{(2\pi)^3}\int
\theta(\omega\cd k)\delta(k^2-\mu^2)d^4k
 \nonumber\\
&&\mbox {\ \ \ \ \ \ \ }\times({\sla p}'-{\sla k}+{\sla
\omega}\tau'+m)\theta[\omega\cd (p'-k)]\nonumber \\
&&\mbox {\ \ \ \ \ \ \ }\times
\delta[(p'-k+\omega\tau')^2-m^2]\frac{d\tau'}{\tau'}
 \nonumber\\
&&\mbox {\ \ \ \ \ \ \ }\times\gamma_{\rho}({\sla p}-{\sla k}+{\sla
\omega}\tau+m)\theta[\omega\cd (p-k)]\nonumber \\
&&\mbox {\ \ \ \ \ \ \ }\times
\delta[(p-k+\omega\tau)^2-m^2]\frac{d\tau}{\tau}
\\
&=&\frac{g^2}{(2\pi)^3}\int
d^2k_{\perp}\int_{\epsilon}^{1-\epsilon}\frac{dx}{2x}\nonumber \\
&&\mbox {\ \ \  }\times
\frac{({\sla p}'-{\sla k}+{\sla \omega}\tau'+m)\gamma_{\rho}({\sla
p}-{\sla k}+{\sla \omega}\tau+m)}{2\omega\cd (p'-k)\tau'\;\;
2\omega\cd (p-k)\tau}.\nonumber
\end{eqnarray}
We have introduced here new integration LF variables ${\bf
k}_{\perp}$ and $x$ in a standard fashion (see Sec.~\ref{sec:LFDSE}).
The singularities of the integrand at $x=0$ and $x=1$ are excluded
by introducing an infinitesimal positive cutoff $\epsilon$. The
values of $\tau$'s are found from the
conservation laws imposed by the delta-functions. Namely,
\begin{subequations}
\label{tautaup}
\begin{multline}
2\omega\cd
(p-k)\tau=m^2-(p-k)^2\\
=(1-x)\left(\frac{{\bf k}_{\perp}^2+m^2}{1-x}+
\frac{{\bf k}_{\perp}^2+\mu^2}{x}-m^2\right),
\label{tau}
\end{multline}
\begin{multline}
2\omega\cd
(p'-k)\tau'=m^2-(p'-k)^2\\
=(1-x')\left(\frac{{{\bf k}'}_{\perp}^2+m^2}{1-x'}+
\frac{{{\bf k}'}_{\perp}^2+\mu^2}{x'}-m^2\right),
\label{taup}
\end{multline}
\end{subequations}
where $x'=\omega\cd k/ \omega\cd p'=x/(1+\alpha)$, ${\bf
k}'_{\perp}={\bf k}_{\perp}-x'{\bg \Delta}$, and ${\bg \Delta}$ is
the part of the three-vector ${\bf q}$, transversal to ${\bg
\omega}$. Note that
\begin{equation}\label{dQ}
{\bg \Delta}^2=Q^2(1+\alpha)-\alpha^2 m^2.
\end{equation}
The formulas~(\ref{tautaup})--(\ref{dQ}) follow from the kinematical
relations listed in Appendix~\ref{kin}.

The contribution of the pair creation diagram, Fig.~\ref{pair},
 is given by
\begin{eqnarray*}
\Gamma_{\rho}^{(pair)}&=&\frac{g^2}{(2\pi)^3}\int
\theta(\omega\cd k)\delta(k^2-\mu^2)d^4k \\
&&\times ({\sla p}'-{\sla k}+{\sla \omega}\tau'+m)\theta[\omega\cd
(p'-k)]\\
&& \times\delta[(p'-k+\omega\tau')^2-m^2]\frac{d\tau'}{\tau'}\\
&&\times\gamma_{\rho}({\sla p}-{\sla k}+{\sla \omega}\tau'-{\sla
\omega}\tau_1+m)\theta[\omega\cd (k-p)]\\
 &&\times\delta[(p-k+\omega\tau'-\omega\tau_1)^2-m^2]
\frac{d\tau_1}{\tau_1}.
\end{eqnarray*}
The line carrying the four-momentum $k_1$ corresponds to an
antifermion and is described by the propagator ${(m-{\sla
k}_1)}\theta(\omega\cd k_1)\delta(k_1^2-m^2)$. It is convenient to
introduce  a  new variable $\tau=\tau'-\tau_1$, instead
of $\tau_1$:
\begin{eqnarray}
 \Gamma_{\rho}^{(pair)}&=&\frac{g^2}{(2\pi)^3}\int
\frac{d^3k}{2\varepsilon_{\bf k}} ({\sla p}'-{\sla k}+{\sla
\omega}\tau'+m) \nonumber \\
&&\times\theta[\omega\cd
(p'-k)]\delta[(p'-k+\omega\tau')^2-m^2]\frac{d\tau'}{\tau'}
 \nonumber\\
&&\times\gamma_{\rho}({\sla p}-{\sla k}+{\sla \omega}\tau+m)
 \theta[\omega\cd (k-p)]\nonumber\\
 &&\times\delta[(p-k+\omega\tau)^2-m^2]
\frac{d\tau}{(\tau'-\tau)}
 \label{2ct2}\\
&=&-\frac{g^2}{(2\pi)^3}\int
d^2k_{\perp}\int_{1+\epsilon}^{1+\alpha-\epsilon}\frac{dx}{2x}\,\nonumber\\
&&\times\frac{({\sla p}'-{\sla k}+{\sla \omega}\tau'+m)\gamma_{\rho}({\sla
p}-{\sla k}+{\sla \omega}\tau+m)}{2\omega\cd (p'-k)\tau'\;\;
2\omega\cd (p-k)(\tau'-\tau)}.\nonumber
\end{eqnarray}
The "shifted" $\tau$ enters the delta-function and is determined
by the same formula~(\ref{tau}) as $\tau$ in other $\Gamma$'s. We
can therefore use the same  kinematics.  The integration over
$d\tau$ by means of the delta-function
\mbox{$\delta[(p-k+\omega\tau)^2-m^2]$} brings the factor
$|\omega\cd (p-k)|$ in the denominator.  We used that
\mbox{$|\omega\cd (p-k)|=-\omega\cd (p-k)$}  inside  the
integration interval $1+\epsilon< x <1+\alpha-\epsilon$.

The contribution  of the left contact term to the EMV is shown in
Fig.~\ref{lrct}(a). The corresponding amplitude has the form
\begin{eqnarray}
\Gamma_{\rho}^{(lct)}&= &\frac{g^2}{(2\pi)^3}\int
\theta(\omega\cd k)\delta(k^2-\mu^2)d^4k\nonumber \\
&&\times({\sla p}'-{\sla k}+{\sla \omega}\tau'+m)\theta[\omega\cd
(p'-k)]\nonumber\\
&&\times\delta[(p'-k+\omega\tau')^2-m^2]\frac{d\tau'}{\tau'}\nonumber \\
&&\times\gamma_{\rho} \left[-\frac{{\sla \omega}}{2\omega\cd (p-k)}\right]
\label{AC1}  \\
&= &-\frac{g^2}{(2\pi)^3}\int
d^2k_{\perp}\nonumber\\
&&\times\int_{\epsilon}^{1+\alpha-\epsilon}\frac{dx}{2x}\,
\frac{({\sla p}'-{\sla k}+{\sla \omega}\tau'+m)\gamma_{\rho}{\sla
\omega}}{2\omega\cd (p'-k)\tau'\;\;2\omega\cd (p-k)}.\nonumber
\end{eqnarray}
For the right contact term, Fig.~\ref{lrct}(b), we get
\begin{eqnarray}
 \Gamma_{\rho}^{(rct)}&= &\frac{g^2}{(2\pi)^3}\int
\theta(\omega\cd k)\delta(k^2-\mu^2)d^4k \nonumber\\
 &&\times\left[-\frac{{\sla \omega}}{2\omega\cd
 (p'-k)}\right]\gamma_{\rho}
({\sla p}-{\sla k}+{\sla \omega}\tau+m)\nonumber\\
&& \times\theta[\omega\cd
(p-k)]\delta[(p-k+\omega\tau)^2-m^2]\frac{d\tau}{\tau}
 \label{AC2}\\
&=&-\frac{g^2}{(2\pi)^3}\int
d^2k_{\perp} \nonumber\\
&&\times\int_{\epsilon}^{1-\epsilon}\frac{dx}{2x}\,
\frac{{\sla \omega}\gamma_{\rho}({\sla p}-{\sla k}+{\sla
\omega}\tau+m)}{2\omega\cd (p'-k)\;\;2\omega\cd (p-k)\tau}.\nonumber
\end{eqnarray}
Finally, the double contact term, Fig.~\ref{double_ct}, yields
\begin{multline}
\label{2ctp} \Gamma_{\rho}^{(2ct)}=\frac{g^2}{(2\pi)^3}\int
d^2k_{\perp}\\
\times\int_{\epsilon}^{+\infty}\frac{dx}{2x}\, \frac{{\sla
\omega}}{2\omega\cd(p'-k)}\,
 \gamma_{\rho}\,
\frac{{\sla \omega}}{2\omega\cd(p-k)}.
\end{multline}

The full EMV is the sum of all the five
contributions~(\ref{GammaA}), (\ref{2ct2})--(\ref{2ctp}). Note
that the limits of integrations over $dx$ are different in these
contributions.  In order to make the calculations easier, we split
the whole region of possible values of $x$ into the three
sub-regions: (1) $\epsilon<x<1-\epsilon$; (2)
$1+\epsilon<x<1+\alpha-\epsilon$; (3)
$1+\alpha+\epsilon<x<+\infty$. We then represent each of the
vertices $\Gamma_{\rho}^{(lct)}$ and $\Gamma_{\rho}^{(2ct)}$ as a
sum of integrals  over  these integration sub-regions, removing
the singularities by means of the same cutoff $\epsilon$:
$$
\Gamma_{\rho}^{(lct)}=\Gamma_{\rho}^{(1,lct)}+\Gamma_{\rho}^{(2,lct)},
$$
where
\begin{eqnarray*}
\Gamma_{\rho}^{(1,lct)}&=&-\frac{g^2}{64\pi^3}\frac{1}{(\omega\cd
p)^2} \int d^2k_{\perp} \\
&&\times\int_{\epsilon}^{1-\epsilon}\frac{dx}{x}\,
\frac{({\sla p}'-{\sla k}+{\sla \omega}\tau'+m)\gamma_{\rho}{\sla
\omega}}{(1+\alpha-x)(1-x)\;\tau'},
\\
\Gamma_{\rho}^{(2,lct)}&=&-\frac{g^2}{64\pi^3}\frac{1}{(\omega\cd
p)^2}\int
d^2k_{\perp}\\
&&\times\int_{1+\epsilon}^{1+\alpha-\epsilon}\frac{dx}{x}\,
\frac{({\sla p}'-{\sla k}+{\sla \omega}\tau'+m)\gamma_{\rho}{\sla
\omega}}{(1+\alpha-x)(1-x)\;\tau'},
\end{eqnarray*}
and, analogously,
$$
\Gamma_{\rho}^{(2ct)}=\Gamma_{\rho}^{(1,2ct)}+\Gamma_{\rho}^{(2,2ct)},
$$
where
\begin{eqnarray*}
\Gamma_{\rho}^{(1,2ct)}&=&\frac{g^2}{64\pi^3}
\frac{\sla{\omega}\gamma_{\rho}\sla{\omega}}{(\omega\cd p)^2}
\int
d^2k_{\perp}\\
&&\mbox{\ \ \ \ \ \ }\times\int_{\epsilon}^{1-\epsilon}
\frac{dx}{x(1+\alpha-x)(1-x)},
\\
\Gamma_{\rho}^{(2,2ct)}&=&\frac{g^2}{64\pi^3}
\frac{\sla{\omega}\gamma_{\rho}\sla{\omega}}{(\omega\cd p)^2}
\int d^2k_{\perp}\\
&&\mbox{\ \ \ \ \ \ }\times\int_{1+\epsilon}^{1+\alpha-\epsilon}
\frac{dx}{x(1+\alpha-x)(1-x)}
\\
&+& \frac{g^2}{64\pi^3}
\frac{\sla{\omega}\gamma_{\rho}\sla{\omega}}{(\omega\cd p)^2}
\int
d^2k_{\perp}\\
&&\mbox{\ \ \ \ \ \ }\times\int_{1+\alpha+\epsilon}^{\infty}
\frac{dx}{x(1+\alpha-x)(1-x)}.
\end{eqnarray*}

After some rearrangement of the contributions, we can represent
the full EMV as
\begin{equation}
\Gamma_{\rho}\equiv\Gamma_{\rho}^{(A)}+\Gamma^{(B)}_{\rho}+\Gamma^{(C)}_{\rho}
\end{equation}
with
\begin{subequations}
\begin{eqnarray}
\Gamma_{\rho}^{(A)}&=&
\Gamma_{\rho}^{(tri)}+\Gamma_{\rho}^{(1,lct)}
+\Gamma_{\rho}^{(rct)}+\Gamma_{\rho}^{(1,2ct)}-\Gamma^{(0)}_{\rho},
\label{GmA}\nonumber \\
\\
\Gamma^{(B)}_{\rho}&=&\Gamma_{\rho}^{(pair)}+
\Gamma_{\rho}^{(2,lct)}- \Gamma_{\rho}^{(1)},
 \label{GmB}\\
\Gamma^{(C)}_{\rho}&=&\Gamma_{\rho}^{(2,2ct)}+
\Gamma_{\rho}^{(0)}+\Gamma_{\rho}^{(1)}.
\label{GmC}
\end{eqnarray}
\end{subequations}

The two new integrals,
\begin{subequations}
 \begin{eqnarray}\label{G0}
\Gamma_{\rho}^{(0)} & = &{\displaystyle  \frac{g^2}{64\pi^3(1+\alpha)}\,
\frac{{\sla \omega} \gamma_{\rho} {\sla \omega}}{(\omega\cd p)^2}\int
d^2k_{\perp}\int_{\epsilon}^{1-\epsilon}\frac{dx}{x}},\\
\Gamma_{\rho}^{(1)} & = & {\displaystyle \frac{g^2}{64\pi^3\alpha}\,
\frac{{\sla \omega} \gamma_{\rho} {\sla
\omega}}{(\omega\cd p)^2}\int
d^2k_{\perp}\int_{1+\epsilon}^{1+\alpha-\epsilon}
\frac{dx}{x(1+\alpha-x)}}\nonumber \\
\label{G1}
\end{eqnarray}
\end{subequations}
have been  introduced in order to make the integrands of
$\Gamma_{\rho}^{(A)}$ and $\Gamma_{\rho}^{(B)}$ non-singular in
$x$, although each  term  on the right-hand sides of
Eqs.~(\ref{GmA}) and~(\ref{GmB}) contains logarithmic divergencies
either at $x=0$ or at $x=1$, or at $x=1+\alpha$.

Taking the sums~(\ref{GmA}) and~(\ref{GmB}), we easily find
\begin{widetext}
\begin{subequations}
\label{GLFT}
\begin{eqnarray}
\label{GLF}
\Gamma_{\rho}^{(A)} & = & {\displaystyle \frac{g^2}{16\pi^3}\int
d^2k_{\perp}\int_0^{1}\frac{dx}{x}\,\left[ \frac{({\sla p}'-{\sla
k}+m)\gamma_{\rho}({\sla p}-{\sla k}+m)}{(\mu^2-2p'\cd
k)(\mu^2-2p\cd k)}-\frac{{\sla \omega} \gamma_{\rho} {\sla
\omega}}{4(1+\alpha)(\omega\cd p)^2} \right]},\\
\Gamma_{\rho}^{(B)} & = & {\displaystyle \frac{g^2}{64\pi^3(\omega\cd p)^2}\int
d^2k_{\perp}\int_1^{1+\alpha}\frac{dx}{x(1+\alpha-x)}}
 {\displaystyle \left[
\frac{({\sla p}'-{\sla k}+{\sla \omega}\tau'+m)\gamma_{\rho}({\sla
p}-{\sla k}+{\sla
\omega}\tau'+m)}{(x-1)\tau'(\tau'-\tau)}-\frac{{\sla \omega}
\gamma_{\rho} {\sla \omega}}{\alpha} \right]}.
\label{GLFB}\end{eqnarray}
\end{subequations}
\end{widetext}
Since  the  regularization in $x$ is no more required,
we set $\epsilon=0$.

The remaining part of the EMV, $\Gamma_{\rho}^{(C)}$, is
given  by a sum of regularized integrals:
\begin{multline}
\Gamma_{\rho}^{(C)}=\frac{g^2}{64\pi^3}\,\frac{{\sla
\omega}\gamma_{\rho}{\sla \omega}}{(\omega\cd
p)^2}\int d^2k_{\perp}
\left[\frac{1}{1+\alpha}\int_{\epsilon}^{1-\epsilon}\frac{dx}{x}\right.\\
+
\frac{1}{\alpha}\int_{1+\epsilon}^{1+\alpha-\epsilon}\frac{dx}{x(1-x)}\\
+
\left.\int_{1+\alpha+\epsilon}^{+\infty}\frac{dx}{x(1+\alpha-x)(1-x)}\right].\\
\label{GLFC}
\end{multline}
%
\subsubsection{Electromagnetic form factors}
We represent the EMV via the following decomposition:
\begin{multline}
\label{dec2} \bar{u}'\Gamma_{\rho}u=\bar{u}'\left\{
{F}_1\gamma_{\rho}+\frac{iF_2}{2m}\sigma_{\rho\nu}q_{\nu}\right.\\
+\left. B_1\left[\frac{{\sla \omega}}{\omega\cd
p}(p+p')_{\rho}-2\gamma_{\rho}\right]+B_2\frac{m\omega_{\rho}}{\omega\cd
p}+B_3\frac{m^2{\sla \omega}\omega_{\rho}}{(\omega\cd
p)^2}\right\}u,
\end{multline}
which is similar, although not identical, to the one used in
Ref.~\cite{km96}. We shall come back to this difference in
Sec.~\ref{discuss}. The decomposition is determined by the
five\footnote{The coincidence of  the number of  form factors with
the number of LF  diagrams  is, of course, by chance.} form
factors $F_{1,2}$, $B_{1-3}$. We shall call  $F_{1,2}$  the
physical form factors, in contrast to the non-physical ones,
$B_{1-3}$, which must be absent in the physically observed EMV.
The  appearance of the three extra form factors is a property of
LFD. Note that they appear in standard non-covariant LFD as well,
but in this approach they can not be separated out from the
physical form factors, and an illusion may occur that the EMV
structure is determined, according to Eq.~(\ref{FFs}), by the two
form factors, as in the Feynman case. This is however not so,
because the electromagnetic current operator in LFD has {\em five}
independent matrix elements, not two. One may argue that even if
each particular  LF  diagram produces a contribution of the
form~(\ref{dec2}) to the EMV, the extra three structures disappear
(i.e., one has $B_{1-3}=0$) after summing up all the
contributions, at least in a given order of perturbation theory.
It would be so if the amplitudes of the  LF  diagrams were given
by convergent integrals, which happens for the pure scalar case
(scalar bosons plus scalar "photon" and no fermions). For systems
involving fermions, however,  the  amplitudes strongly diverge.
Their regularization, as will be shown below, may give rise to the
 appearance of extra ($\omega$-dependent) spin structures in
the  perturbative and non-perturbative  regularized EMV's. In this
sense  the situation is  analogous to the case of the fermion
self-energy discussed in the previous section.

The procedure to calculate the form factors is quite similar to that
exposed in Ref.~\cite{km96}. We define the matrix
\begin{equation}\label{OO}
 O_{\rho}=\frac{({\sla p}'+m)\Gamma_{\rho}({\sla
p}+m)}{4m^2}
 \end{equation}
and the following contractions:
\begin{eqnarray}
c_1&=&\mbox{Tr}\left\{O_{\rho}\gamma^{\rho}\right\},\nonumber \\
c_2&=&\frac{iq_{\nu}}{2m}\mbox{Tr}
\left\{O_{\rho}\sigma^{\rho\nu}\right\}, \nonumber \\
c_3&=&\frac{(p+p')^{\rho}}{\omega\cd p}\mbox{Tr}\left\{O_{\rho}{\sla
\omega}\right\},\label{contrc}\\
c_4&=&\frac{m\omega^{\rho}}{\omega\cd
p}\mbox{Tr}\left\{O_{\rho}\right\}, \nonumber \\
c_5&=&\frac{m^2\omega^{\rho}}{(\omega\cd
p)^2}\mbox{Tr}\left\{O_{\rho}{\sla \omega}\right\}.\nonumber
\end{eqnarray}
Now, using Eq.~(\ref{dec2}), we can get the following relations
between the form factors and the quantities $c_{1-5}$:
\begin{eqnarray}
c_1 & = & (2-4\eta) F_1-6\eta F_2 +
2(4\eta+\alpha) B_1\nonumber \\
&&+(2+\alpha) B_2+2(1+\alpha) B_3,\nonumber \\
 c_2 & = &  6\eta F_1+2\eta(2-\eta)
F_2-2\eta(4-\alpha) B_1\nonumber \\
&&+\eta(2+\alpha)
B_2+\frac{\alpha^2}{2} B_3,\nonumber \\
c_3 & = & 2(2+\alpha) F_1-2\eta(2+\alpha)
F_2\nonumber \\
&&+4[2\eta(1+\alpha)+\alpha] B_1+(2+\alpha)^2
B_2\nonumber \\
&&+2(2+\alpha)(1+\alpha) B_3,\nonumber\\
c_4 & = & (2+\alpha)\left[ F_1-\eta F_2+\alpha
B_1\right],\nonumber \\
c_5 & = &  2(1+\alpha) F_1-
\frac{\alpha^2}{2} F_2+2\alpha(1+\alpha) B_1,\label{eqc5}
\end{eqnarray}
where $\eta=Q^2/(4m^2)$. Solving the system of linear
equations~(\ref{eqc5}) with respect to $F_{1,2}$  and $B_{1-3}$,
we express  them  through $c_{1-5}$. These expressions are lengthy
and we do not give them here. Then, substituting the
formulas~(\ref{GLFT}), (\ref{GLFC}) for $\Gamma^{(A)}_{\rho}$,
$\Gamma^{(B)}_{\rho}$, and $\Gamma_{\rho}^{(C)}$ into
Eq.~(\ref{OO}), calculating the traces~(\ref{contrc}), and
expressing the scalar products of the four-vectors $p$, $p'$, $k$,
and $\omega$ through the two-dimensional vectors ${\bf
k}_{\perp}$, ${\bg \Delta}$, and the scalars $x$, $\alpha$, we
cast  each form factor in the form of a three-dimensional integral
over $d^2k_{\perp}dx$. The expressions for  the  scalar products
through the integration variables and the momentum transfer are
given in Appendix~\ref{kin}.
\subsubsection{Regularization with rotationally non-invariant cutoffs}
We give here the final expressions for the form factors found by
using the rotationally non-invariant cutoffs $\Lambda_{\perp}$ and
$\epsilon$ imposed on the variables $|{\bf k}_{\perp}|$  and $x$.
As we said above, the vertex functions $\Gamma_{\rho}^{(A)}$ and
$\Gamma_{\rho}^{(B)}$ do not require to introduce a cutoff in $x$.
As far as $\Gamma_{\rho}^{(C)}$ is concerned, it is represented by
a sum of integrals, each requiring regularization and thus
depending on $\epsilon$. However, as can be established by direct
integration in Eq.~(\ref{GLFC}), the sum itself has  a finite
limit when $\epsilon\to 0$:
\begin{equation}
\label{GLFC1}
\Gamma_{\rho}^{(C)}=\frac{g^2\Lambda_{\perp}^2}{32\pi^2}\,
\frac{\log(1+\alpha)}{\alpha(1+\alpha)}\,\frac{{\sla
\omega}\omega_{\rho}}{(\omega\cd p)^2}.
\end{equation}
We therefore remain with the cutoff $\Lambda_{\perp}$ only. Note
that the mutual cancellation of  the  terms singular in $x$, which
happens for the full EMV, is connected with rather weak
(logarithmic) divergence of the corresponding integrals, so that
the cutoff $\Lambda_{\perp}$ is enough to make them finite. The
same took place in the fermion self-energy case (see
Sec.~\ref{regLd}). As we shall see below [Eq.~(\ref{aL1B}) in
Appendix~\ref{ffs}], the quadratically divergent term $\sim
\sla{\omega}\omega_{\rho}\Lambda_{\perp}^2$ results also from
$\Gamma^{(B)}_{\rho}$ (but not from $\Gamma^{(A)}_{\rho}$) and
cancels in  the sum with $\Gamma_{\rho}^{(C)}$.

The details of calculations of all the contributions $(A,B,C)$ are
given in Appendix~\ref{ffs}. Adding all them together, we arrive
at the following final result for the form factors:
\begin{subequations}
\label{FTnoncov}
\begin{eqnarray}
\label{F1noncov} F_1 & = &
\frac{g^2}{16\pi^2}\log\frac{\Lambda_{\perp}}{m}+
\frac{g^2}{4\pi^2}\left(\log\frac{m}{\mu}-\frac{7}{8}\right)\nonumber\\
&&-\frac{g^2Q^2}{24\pi^2m^2}\left(\log\frac{m}{\mu}-\frac{9}{8}\right)+O(Q^4),
\\
\label{F2noncov} F_2 & = &  \frac{3g^2}{16\pi^2}
-\frac{g^2Q^2}{32\pi^2m^2}+O(Q^4),
\\
\label{B1noncov} B_1 & = &  -\frac{g^2}{64\pi^2},
\\
\label{B2noncov} B_2 & = &  -\frac{g^2}{32\pi^2},
\\
\label{B3noncov} B_3 & = &
-\frac{g^2(m^2-\mu^2)}{16\pi^2m^2}\log\frac{\Lambda_{\perp}}{m}\nonumber\\
&&+\frac{g^2}{32\pi^2m^2}\left[\frac{Q^2}{\alpha}+\mu^2\left(
2\log\frac{m}{\mu}+1\right)\right].
\end{eqnarray}
\end{subequations}
Note that the terms $\sim\log\frac{\Lambda_{\perp}}{m}$ cancel in
the form factors $B_1$ and $B_2$. These form factors are finite
and do not depend on any cutoff, in contrast to $B_3$ which does
not have a finite limit when $\Lambda_{\perp}\to \infty$. As we
have already mentioned, neither of the form factors depend on the
cutoff $\epsilon$. At the same time, the amplitude of each of the
LF diagrams shown in Figs.~\ref{triangle}--\ref{double_ct},
diverges like $\Lambda_{\perp}^2\log\epsilon$. It means that the
senior divergent terms $\sim\Lambda_{\perp}^2\log\epsilon$ and
$\sim\Lambda_{\perp}^2$ cancel after  the  incorporation of  all
the LF diagrams. However, the form factors $B_{1-3}$ which must be
absent in the physical EMV remain non-zero values. The same
happened for the $\sim \sla{\omega}$ term in the full self-energy,
Eq.~(\ref{Sigdecomp}).

We also  see that $B_3$ has  a  pole at $\alpha=0$. One cannot set
here $Q^2=0$, since it is impossible to keep fixed
$\alpha=(\omega\cd q) /(\omega\cd p)$ when $q\to 0$. More
precisely, since ${\bg \Delta}^2\ge 0$, from Eq.~(\ref{dQ}) we
have $Q^2\ge \frac{\alpha^2 m^2}{1+\alpha}$. However, nothing
prevents us to take $\alpha\to 0$ at fixed $Q^2$. As  the
inspection shows, the singular term $\sim 1/\alpha$ in $B_3$
results from $\Gamma^{(pair)}_{\rho}$,   Eq.~(\ref{2ct2}).

The standard renormalization procedure~(\ref{FFren}) affects the
form factor $F_1$ only, so that
\begin{equation}
\label{F1ren}
F_1^{ren}=1-\frac{g^2Q^2}{24\pi^2m^2}\left(\log\frac{m}{\mu}-\frac{9}{8}\right)+O(Q^4).
\end{equation}
The cutoff-dependent term $\sim\log(\Lambda_{\perp}/m)$ cancels,
as it should. The renormalized form factor $F_2^{ren}$ coincides
with $F_2$, Eq.~(\ref{F2noncov}). It is important to note that in
order to eliminate the three extra form factors $B_{1-3}$, one
should introduce  new $\omega$-dependent counterterms into the
interaction Hamiltonian. Since $B_3$ non-trivially depends on $q$
through $Q^2=-q^2$ and $\alpha=\omega\cd q/\omega\cd p$, its
decomposition in powers of $q$ contains an infinite number of
terms. Therefore, to eliminate $B_3$ we need an infinite number of
local counterterms, for this particular type of regularization.
%
\subsubsection{\label{PVLFD} Invariant Pauli-Villars regularization}
We can now calculate the form factors within LFD, but using the
invariant PV regularization instead of the LF cutoffs
$\Lambda_{\perp}$ and $\epsilon$.

\vspace{0.1cm} \underline{Regularization by  a  PV
boson.} We start with the regularization by one PV boson only.
Since each of the  LF  diagrams contains one bosonic
propagator, this regularization reduces to a simple subtraction
\begin{equation*}
{\cal F}^{PV,\,b}={\cal F}(Q^2,m,\mu)-{\cal F}(Q^2,m,\mu_1).
\end{equation*}
 As before, ${\cal F}$ denotes any of the form factors $F_i$
or $B_i$.  In the limit of large $\mu_1$ we have
\begin{eqnarray*}
F_1(Q^2,m,\mu_1)&=&\frac{g^2}{16\pi^2}\log\frac{\Lambda_{\perp}}{m}-\frac{g^2}{16\pi^2}
\left(\log\frac{\mu_1}{m}-\frac{1}{4}\right),\\
F_2(Q^2,m,\mu_1)&=&0,
\end{eqnarray*}
for arbitrary finite $Q^2$. Subtracting these expressions from
those given by Eqs.~(\ref{F1noncov}) and~(\ref{F2noncov}),
respectively, we get
\begin{subequations}
\label{F12TPV}
\begin{eqnarray}
\label{F1PV} F_1^{PV,\,b} & = &
\frac{g^2}{16\pi^2}\log\frac{\mu_1}{m}+
\frac{g^2}{4\pi^2}\left(\log\frac{m}{\mu}-\frac{15}{16}\right)\nonumber\\
&&-\frac{g^2Q^2}{24\pi^2m^2}\left(\log\frac{m}{\mu}-\frac{9}{8}\right)+O(Q^4),
\\
\label{F2PV} F_2^{PV,\,b} & = &
\frac{3g^2}{16\pi^2} -\frac{g^2Q^2}{32\pi^2m^2}+O(Q^4).
\end{eqnarray}
\end{subequations}
In spite of the fact that Eq.~(\ref{F1PV}) differs from
Eq.~(\ref{F1noncov}), both equations lead to the same renormalized
form factor $F_1^{ren}$, Eq.~(\ref{F1ren}). The renormalized
physical form factors  obtained  for the invariant and
non-invariant types of  regularization do therefore
coincide.

Concerning the non-physical form factors, the situation is quite
different. Since $B_{1-3}$ were calculated for arbitrary $\mu$, we
find from Eqs.~(\ref{B1noncov})--(\ref{B3noncov}):
\begin{subequations}
\label{BTPV}
\begin{equation}
\label{B12PV} B_1^{PV,\,b}=B_2^{PV,\,b}=0,
\end{equation}
\begin{multline}
\label{B3PV}
B_3^{PV,\,b}=-\frac{g^2(\mu_1^2-\mu^2)}{16\pi^2m^2(1+\alpha)}\log
\frac{\Lambda_{\perp}}{m}\\
-\frac{g^2}{32\pi^2m^2}\left(\mu_1^2-\mu^2-2\mu^2\log\frac{m}{\mu}-
2\mu_1^2\log\frac{\mu_1}{m}\right).\
\end{multline}
\end{subequations}
The formulas~(\ref{F12TPV}) are valid for $\mu\to
0$ and $\mu_1\to\infty$, while Eqs.~(\ref{BTPV})
hold for arbitrary $\mu$ and $\mu_1$. The non-physical form
factors $B_{1,2}$ turned into zero, while $B_3^{PV,\,b}\neq 0$
and still diverges logarithmically.

\vspace{0.1cm} \underline{Regularization by  one  PV boson plus
one  PV fermion.} Since the PV regularization by one boson only
does not cancel the form factor $B_3$, we have to introduce  in
addition a  PV fermion. In that case, no contact terms appear at
all, and the only diagrams which contribute to the EMV are the
triangle (Fig.~\ref{triangle}) and pair creation (Fig.~\ref{pair})
diagrams. Each of them contains two fermion propagators, both
being subject to the PV regularization. For this reason we need to
know the vertex with different internal fermions. We denote the
masses of these fermions by $m_i$ and $m_{i'}$, with the indices
$i$ and $i'$ being either 0 or 1.  Let $m_0\equiv m$ and $m_1$ be
the physical and PV fermion masses, respectively. Analogously, we
denote the physical and PV boson masses by $\mu_0\equiv \mu$ and
$\mu_1$.

The PV regularized EMV is defined by
$$
\Gamma_{\rho}^{PV,\,b+f}=\sum_{i,i',j=0}^1
(-1)^{i+i'+j}\Gamma_{\rho}(m_i,m_{i'},\mu_j),
$$
where  the notation  $\Gamma_{\rho}(m_i,m_{i'},\mu_j)$ stands for
the corresponding initial EMV calculated for the internal
particles with the masses $m_i$, $m_{i'}$, and $\mu_j$. The
quantity $\Gamma_{\rho}(m_i,m_{i'},\mu_j)$ itself is given by a
sum of Eqs.~(\ref{GLFT}), changing everywhere
\begin{eqnarray*}
{\sla p}-{\sla k}+m&\to &{\sla p}-{\sla k}+m_i,\\
{\sla p}'-{\sla k}+m&\to& {\sla p}'-{\sla k}+m_{i'},\\
\tau&\to &\frac{m_i^2-(p-k)^2}{2\omega\cdot
(p-k)},\\
\tau'&\to&
\frac{m_{i'}^2-(p'-k)^2}{2\omega\cdot (p'-k)},
\end{eqnarray*}
and $\mu\to\mu_j$. The vertex $\Gamma_{\rho}^{(C)}$,
Eq.~(\ref{GLFC}), turns into zero already by the boson PV
regularization and does not contribute to the PV regularized form
factors.

It can be shown that we do not need to introduce  the  PV fermion
to regularize the form factors $F_{1,2}$ and $B_{1,2}$. In other
words, in the limit $m_1\to\infty$ we  would obtain the same
formulas~(\ref{F12TPV})--(\ref{B12PV}) as without the  PV fermion
at all. So, the only thing to do is to calculate $B_3^{PV,\,b+f}$.
This calculation is in principle quite similar to that performed
above, but  the  algebra is more lengthy. We represent
$B_3^{PV,\,b+f}$ as follows:
\begin{equation}
\label{B3PVbf}
B_3^{PV,\,b+f}=\sum_{i,i'=0}^1(-1)^{i+i'}\sum_{N=A,B}\,
B_3^{(N)PV,\,b}(m_i,m_{i'}),
\end{equation}
where
\begin{equation}
\label{B3PVb} B_3^{(N)PV,\,b}(m_i,m_{i'})=\sum_{j=0}^1
(-1)^jB_3^{(N)}(m_i,m_{i'},\mu_j)
\end{equation}
and $B_3^{(N)}(m_i,m_{i'},\mu_j)$ is found through the vertex
$\Gamma_{\rho}^{(N)}(m_i,m_{i'},\mu_j)$. Though the order of
summations in Eqs.~(\ref{B3PVbf}) and~(\ref{B3PVb}) does not
matter, it is convenient to calculate first
$B_3^{(N)PV,\,b}(m_i,m_{i'})$, separately for the two vertices,
then add the results, and finally sum over the PV fermion indices.
Omitting rather tiresome manipulations, we give here  the
expressions for the functions $B_3^{(N)PV,\,b}(m_i,m_{i'})$, in
order to see  the details of the  cancellations:
\begin{widetext}
\begin{subequations}
\label{B3ABPV}
\begin{eqnarray}
B_3^{(A)PV,\,b}(m_i,m_{i'}) & = & {\displaystyle
-\frac{g^2(\mu_1^2-\mu^2)(2+\alpha)}{16\pi^2m^2(1+\alpha)^2}
\log\frac{\Lambda_{\perp}}{m}}\nonumber\\
&  & + {\displaystyle \frac{g^2}{32\pi^2m^2}\left[2\mu_1^2
\log\frac{\mu_1}{m}+2\mu^2\log\frac{m}{\mu}-3(\mu_1^2-\mu^2)\right]+
R(m_i,m_{i'})},\label{B3APVm1m2} \\
\nonumber\\
 B_3^{(B)PV,\,b}(m_i,m_{i'}) & = & {\displaystyle
\frac{g^2(\mu_1^2-\mu^2)}{16\pi^2m^2(1+\alpha)^2}\log\frac{\Lambda_{\perp}}{m}
+\frac{g^2(\mu_1^2-\mu^2)}{16\pi^2m^2}-R(m_i,m_{i'})},\label{B3BPVm1m2}
\end{eqnarray}
\end{subequations}
where
%
\begin{equation*}
R(m_i,m_{i'})=-\frac{g^2(\mu_1^2-\mu^2)}{64\pi^2m^2Q^2}\left\{
(m_i^2-m_{i'}^2)\log\frac{m_i^2}{m_{i'}^2}
-2Q^2\log\frac{m_im_{i'}}{m^2} +s_Q\log\left[
\frac{(m_i^2-m_{i'}^2)^2-(Q^2-s_Q)^2}{(m_i^2-m_{i'}^2)^2-(Q^2+s_Q)^2}\right]\right\}
\end{equation*}
\end{widetext}
and $s_Q=\sqrt{m_i^4-2m_i^2(m_{i'}^2-Q^2)+(m_{i'}^2+Q^2)^2}$. Both
contributions~(\ref{B3ABPV}) are rather
complicated functions depending on the physical and PV masses, as
well as on $Q^2$. In their sum, however, the dependence on the
fermion mass $m_1$ and  on  $Q^2$ drops out completely,
and for $\sum_{N=A,B}B_3^{(N)PV,\,b}(m_i,m_{i'})$ we arrive at the
same expression~(\ref{B3PV}) found previously without any fermion
PV regularization.  The  final summation over the
fermion PV indices, as prescribed by Eq.~(\ref{B3PVbf}), turns
$B_3$ into zero:
\begin{equation}
\label{B3PVfin} B_3^{PV,\,b+f}=0.
\end{equation}

We can thus formulate the main results of this section: the full
cancellation of the three non-physical form factors is achieved by
the double  (i.e., bosonic + fermionic)  PV regularization,
whereas the single boson PV regularization cancels only two of the
three non-physical form factors. The renormalized physical form
factors obtained by using the PV regularization (both single and
double) coincide with those calculated through the rotationally
non-invariant  cutoffs.
%
\subsection{\label{EMFey} Calculation in the four-dimensional \\
Feynman approach}
The amplitude of the Feynman triangle diagram which determines the
EMV in the given order of perturbation theory is
\begin{multline}
 \label{eq1}
\Gamma_{\rho}=\frac{ig^2}{(2\pi)^4}\int d^4k\frac{1}{[k^2-\mu^2+i0]}\\
 \times\frac{({\sla
p}'-{\sla k}+m)\gamma_{\rho}({\sla p}-{\sla
k}+m)}{[(p-k)^2-m^2+i0][(p'-k)^2-m^2+i0]},
\end{multline}
where $p^2=p'^2=m^2$. As we shall see,  the spin structure  of the
Feynman EMV depends on how it is regularized, in contrast to the
self-energy, Eq.~(\ref{sef}). In order to avoid overloading  the
formulas by additional indices indicating the type of  the
cutoffs, we use in this section the same notations  for the
Feynman EMV and  the  form factors as for the LFD ones.
\subsubsection{\label{GFn} Regularization with rotationally
non-invariant cutoffs}
 As we did for the self-energy, Sec.~\ref{Sigma_Fey}, we first
find the form factors by integrating the Feynman
amplitude~(\ref{eq1}) in terms of the LF variables restricted by
the transverse and longitudinal cutoffs. We shall see that such a
procedure generates extra (non-physical) form factors $B_{1-3}$
[see Eq.~(\ref{dec2})], in a very similar way as the use of the
LFD rules.

Rewriting the integrand in Eq.~(\ref{eq1}) through the plus-,
minus-, and transverse components of the four-vector $k$, we get
\begin{eqnarray}
\Gamma_{\rho}&&=\frac{ig^2}{32\pi^4}\int d^2k_{\perp}dk_+dk_-Ê\nonumber\\
&&\times\frac{({\sla p}'-{\sla k}+m)\gamma_{\rho}}{[(p'_+-k_+)(p'_- -k_-)-
({\bf p}'_{\perp}-{\bf k}_{\perp})^2-m^2+i0]}  \nonumber\\
&& \times\frac{({\sla p}-{\sla k}+m)}
{[(p_+-k_+)(p_--k_-)- ({\bf p}_{\perp}-{\bf
k}_{\perp})^2-m^2+i0]}\nonumber\\
&&\times\frac{1}{[k_+k_--{\bf k}_{\perp}^2-\mu^2+i0]}.
\label{eq2}\end{eqnarray}
We take the reference frame where ${\bf p}=0$. We then choose
$p_+=p'_+$. The latter condition is equivalent, in the language of
covariant LFD, to $\alpha=(\omega\cd q)/(\omega\cd p)=0$. In our
LFD calculation of the form factors in Sec.~\ref{EMvLFD}, we
initially kept $\alpha$ to be non-zero and went over to the limit
$\alpha\to 0$ after summing up all the  LF contributions. A smooth
limit $\alpha\to 0$ was important in the LFD framework, since  the
amplitudes  of different diagrams shown in
Figs.~\ref{triangle}--\ref{double_ct} depend  on $\alpha$. The
Feynman amplitude~(\ref{eq1}) is well-defined for any values of
its arguments $p$ and $p'$, in particular, for $p_+=p'_+$. We can
therefore safely  set  $p_+ -p'_+=0$ from the very beginning.

Similarly to Eq.~(\ref{sigmap}) for the self-energy, we represent
the full EMV as a sum of the two terms:
\begin{equation}
\label{2t} \Gamma_{\rho}\equiv
\Gamma_{\rho}^{(p+a)}+\Gamma_{\rho}^{(zm)},
\end{equation}
corresponding to the "normal" contribution and  the zero modes,
respectively.

Like the self-energy case, the "normal" part
$\Gamma_{\rho}^{(p+a)}$ is determined by the sum of the pole and
arc contributions. It is calculated in Appendix \ref{derEMV},
Sec.~\ref{parcEMV}. The result reads
\begin{eqnarray}
\Gamma_{\rho}^{(p+a)} &=&\frac{g^2}{16\pi^3}\int
d^2k_{\perp}\int_{0}^{1}\frac{dx}{x}\label{GLF1}
\\
&&\times\left[ \frac{({\sla p}'-{\sla k}+m)\gamma_{\rho}({\sla
p}-{\sla k}+m)}{(\mu^2-2p'\cd k)(\mu^2-2p\cd
k)}-\frac{\sla{\omega}\gamma_{\rho}\sla{\omega}}{4(\omega\cd p)^2}
\right].\nonumber 
\end{eqnarray}
Comparing the right-hand sides of Eq.~(\ref{GLF1}) for
$\Gamma_{\rho}^{(p+a)}$ and Eq.~(\ref{GLF}) for the part
$\Gamma_{\rho}^{(A)}$ of the EMV calculated within LFD, we see
that they exactly coincide with each other at $\alpha=0$:
\begin{equation}
\label{GpaLF}
\Gamma_{\rho}^{(p+a)}=\left.\Gamma_{\rho}^{(A)}\right|_{\alpha=0}.
\end{equation}
We have already encountered a similar situation above, in
Sec.~\ref{Sigma_Fey}, where it was shown that the pole plus arc
contribution to the Feynman expression for the self-energy exactly
coincided with the full LFD self-energy [see Eq.~(\ref{SigmaA2})].
Evidently, the vertex $\Gamma_{\rho}^{(p+a)}$ can be also
represented via form factors, in the form of the
decomposition~(\ref{dec2}), but with its "own" form factors
$F_{1,2}^{(p+a)}$, $B_{1-3}^{(p+a)}$. Form factors for
$\Gamma_{\rho}^{(A)}$ were found in Appendix~\ref{ffs}. Due to the
identity~(\ref{GpaLF}), in order to derive the form factors
$F_{1,2}^{(p+a)}$ and $B_{1-3}^{(p+a)}$, we can make use of
Eq.~(\ref{FFc}) with the coefficients from
Eqs.~(\ref{aL1A})--(\ref{aregA}), taken at $\alpha=0$. We thus
find
\begin{subequations}
\label{BTFLFA}
\begin{eqnarray}
F_1^{(p+a)}&=&\frac{g^2}{16\pi^2}\log\frac{\Lambda_{\perp}}{m}+\frac{g^2}{4\pi^2}
\left(\log\frac{m}{\mu}-\frac{7}{8}\right)\nonumber\\
\label{F1FLFA}&&
+\frac{g^2Q^2}{24\pi^2m^2}\left(\log\frac{m}{\mu}-\frac{9}{8}\right)+O(Q^4),\\
\label{F2FLFA}
F_2^{(p+a)}&=&\frac{3g^2}{16\pi^2}-\frac{g^2Q^2}{32\pi^2m^2}+O(Q^4),\\
\label{B1FLFA}
 B_1^{(p+a)}&=&\phantom{-}\frac{g^2}{16\pi^2}
 \,\log\frac{\Lambda_{\perp}}{m}-\frac{g^2}{64\pi^2}
 +\frac{g^2}{16\pi^2}\varphi(Q^2),\nonumber\\
 &&\\
\label{B2FLFA}
 B_2^{(p+a)}&=&-\frac{g^2}{8\pi^2}
 \,\log\frac{\Lambda_{\perp}}{m}-\frac{g^2}{32\pi^2}
 -\frac{g^2}{8\pi^2}\varphi(Q^2),\\
 B_3^{(p+a)}&=&-\frac{g^2}{32\pi^2m^2}\left(6m^2-4\mu^2+Q^2\right)
 \log\frac{\Lambda_{\perp}}{m}
\nonumber\\
 & & +\frac{g^2}{32\pi^2m^2}\,\left[
 \mu^2\left(2\log\frac{m}{\mu}+1\right)\right.\nonumber\\
 &&\left.
- (4m^2-2\mu^2+Q^2)\varphi(Q^2)\right],
 \label{B3FLFA}
\end{eqnarray}
\end{subequations}
where the function $\varphi(Q^2)$ is given by Eq.~(\ref{funphi}).
Note that $\varphi(0)=0$. As before, the form factors
$F_{1,2}^{(p+a)}$ are calculated for $\mu\to 0$ and decomposed in
powers of $Q^2$, whereas $B^{(p+a)}_{1-3}$ are exact in this
sense.

The zero-mode contribution $\Gamma_{\rho}^{(zm)}$ is calculated in
Appendix~\ref{derEMV}, Sec.~\ref{zmEMV}. The expression for
$\Gamma_{\rho}^{(zm)}$ is given by Eq.~(\ref{eq19}). We represent
it in the form of decomposition~(\ref{dec2}). Then the form
factors in this decomposition have the form

\begin{subequations}
\label{BTFLFZM}
\begin{eqnarray}
\label{F1FLFZM} F_{1}^{(zm)} & = & 0,\\
\label{F2FLFZM} F_{2}^{(zm)} & = & 0,\\
 \label{B1FLFZM}
 B_1^{(zm)}&=&-\frac{g^2}{16\pi^2} \,\log\frac{\Lambda_{\perp}}{m}
 +\frac{g^2}{16\pi^2}\varphi(Q^2),\\
\label{B2FLFZM}
 B_2^{(zm)}&=&\frac{g^2}{8\pi^2}\,\log\frac{\Lambda_{\perp}}{m}
 +\frac{g^2}{8\pi^2}\varphi(Q^2),\\
 B_3^{(zm)}&=&\frac{g^2}{32\pi^2m^2}\left(6m^2-4\mu^2+Q^2\right)
 \log\frac{\Lambda_{\perp}}{m}
+\frac{g^2}{32\pi^2}
\nonumber\\
 &&-\frac{g^2}{32\pi^2m^2}\,\left\{
 \mu^2\left(2\log\frac{m}{\mu}+1\right)\right.\nonumber\\
 &&\left. \phantom{\int}- (4m^2-2\mu^2+Q^2)\varphi(Q^2)\right\}. \label{B3FLFZM}
\end{eqnarray}
\end{subequations}

The final expressions for the form factors are given by the sum of
the corresponding quantities from Eqs.~(\ref{BTFLFA})
and~(\ref{BTFLFZM}):
\begin{subequations}
\label{BTFLF}
\begin{eqnarray}
\label{F1FLF}
F_1&=&\frac{g^2}{16\pi^2}\log\frac{\Lambda_{\perp}}{m}+
\frac{g^2}{4\pi^2}\left(\log\frac{m}{\mu}-\frac{7}{8}\right)\nonumber\\
&&-\frac{g^2Q^2}{24\pi^2m^2}\left(\log\frac{m}{\mu}-\frac{9}{8}\right)
+O(Q^4),\\
\label{F2FLF} F_2&=&\frac{3g^2}{16\pi^2}-\frac{g^2Q^2}{32\pi^2m^2}+O(Q^4),\\
\label{B1FLF}
 B_1&=&-\frac{g^2}{64\pi^2},\\
\label{B2FLF}
 B_2&=&-\frac{g^2}{32\pi^2},\\
 B_3&=&\phantom{-}\frac{g^2}{32\pi^2}.
 \label{B3FLF}
\end{eqnarray}
\end{subequations}
The expressions for $B_{1-3}$ are exact, i.e.,  these formulas are
valid for any $Q^2$ and $\mu$. That is, the $Q^2$-dependence
coming from $B^{(p+a)}_{1-3}$,
Eqs.~(\ref{B1FLFA})--(\ref{B3FLFA}), as well as the
$\Lambda_{\perp}$-dependent terms are exactly canceled by the
corresponding zero mode contributions $B^{(zm)}_{1-3}$.  The form
factors $F_{1,2}$ and $B_{1,2}$  are the same as those obtained
from  the  LFD diagrammatic approach, in Eqs.~(\ref{FTnoncov}),
while $B_3$'s in Eqs.~(\ref{B3noncov}) and~(\ref{B3FLF}), are
different. Moreover $B_{1-3}$ are not zeroes, though we started
from the Feynman expression for the EMV, which initially had no
any relation to the light front (i.e., to the  four-vector
$\omega$). We will discuss this situation in more detail below, in
Sec.~\ref{discuss}.
%
\subsubsection{\label{PVFey} Invariant regularization}
To complete our analysis, we give here the form factors obtained
from Eq.~(\ref{eq1}) by using  an  invariant regularization of the
divergent integral over $d^4k$. The standard recipe consists in
using the Feynman parametrization, then making the Wick rotation
and imposing a cutoff $\Lambda^2$ on the modulus of the Euclidean
four-momentum  squared, similarly to how it has been done in
Sec.~\ref{ir}.  After that, identifying the structure of the
EMV~(\ref{eq1}) with the decomposition~(\ref{FFs}), we reproduce
the well-known result:\footnote{$F_2(Q^2=0)$ is an anomalous
magnetic moment. Introducing the coupling constant
$\alpha=g^2/(4\pi)$, we get in the Yukawa model:
$F_2(0)=3\alpha/(4\pi)$ that differs by a factor of $3/2$ from the
QED value $\alpha/(2\pi)$.}
\begin{subequations}
\label{F12fey}
\begin{eqnarray}
F_1&=&\frac{g^2}{32\pi^2}\log\frac{\Lambda^2}{m^2}+
\frac{g^2}{4\pi^2}\left(\log\frac{m}{\mu}-\frac{15}{16}\right)\nonumber \\
&&-\frac{g^2Q^2}{24\pi^2m^2}\left(\log\frac{m}{\mu}-\frac{9}{8}\right)+O(Q^4),
\label{F1fey}\\
F_2&=&\frac{3g^2}{16\pi^2}-\frac{g^2Q^2}{32\pi^2m^2}+O(Q^4).
\label{F2fey}
\end{eqnarray}
\end{subequations}
Eq.~(\ref{F2fey}) exactly coincides with Eq.~(\ref{F2FLF})
obtained for the rotationally non-invariant cutoffs. The
formulas~(\ref{F1fey}) and~(\ref{F1FLF}) differ by a constant
(i.e., $Q^2$-independent) part. The corresponding form factors
$F_1^{ren}$ however also coincide after the
renormalization~(\ref{FFren}).

In principle, one might identify the structure of the Feynman EMV
(regularized by means of an invariant cutoff!) with the LF
decomposition~(\ref{dec2}) containing five form factors. In this
case  the formulas~(\ref{F12fey}) would be found for $F_{1,2}$,
while  one would arrive at  the evident result $B_{1-3}=0$ for the
other form factors.

The  regularization of the EMV~(\ref{eq1}) by means of one PV
boson results in the same formulas~(\ref{F12fey}), changing
$\Lambda^2$ to $\mu_1^2$.
%
\section{\label{discuss} Discussion}
We have shown above that the perturbative spin-1/2 fermion
self-energy and the EMV calculated in LFD depend on the
orientation of the LF plane, when  the  traditional regularization
in terms  of the transverse and longitudinal cutoffs is applied to
the corresponding amplitudes. This dependence reveals itself in
the appearance of extra spin structures in  the decompositions of
the self-energy and  the  EMV in invariant amplitudes. In
covariant LFD on the plane $\omega\cd x=0$ with $\omega^2=0$, it
is conveniently parameterized through the four-vector $\omega$.
The corresponding decompositions for the self-energy and  the  EMV
are given by Eqs.~(\ref{Sigdecomp}) and~(\ref{dec2}),
respectively. The structure of the  LF self-energy is
characterized by the three scalar functions ${\cal A}$, ${\cal
B}$, and $\tilde{C}={\cal C}(p^2)+C_{fc}$, instead of the two
usual ones, while the on-energy-shell EMV contains five form
factors, the two standard $F_{1,2}$ and the three extra,
$B_{1-3}$, ones.

Performing calculations of the self-energy by integrating the
Feynman amplitude written in terms of the LF variables with the
same cutoffs as in LFD, we did not encounter any
$\omega$-dependence in the final result, Eq.~(\ref{sigmafin}). The
latter therefore does not coincide with the LFD
self-energy~(\ref{Sigdecomp}). For  the  EMV, both methods lead to
$\omega$-dependent structures with  five form factors. However,
the EMV calculation by means of the LFD rules results in the set
of form factors~(\ref{FTnoncov}), whereas the calculation of the
Feynman amplitude in terms of the LF variables gives a different
set of form factors~(\ref{BTFLF}). Though this difference concerns
the form factor $B_3$ only, the corresponding EMV's do not
coincide nevertheless with each other.

The  extra  form factors do not, of course, appear when the
Feynman amplitude is calculated with a spherically symmetric
cutoff in four-dimensional space. This can be achieved, e.g., by
imposing a direct cutoff on the Wick rotated  integration variable
in Euclidean space $|k^2|={\bf k}^2+k_4^2<\Lambda^2$  or by using
the PV regularization. In the case of LFD,  we deal with
three-dimensional integration variables. Therefore,  constructing
rotationally invariant (i.e., $\omega$-independent) cutoffs  which
restrict a three-dimensional integration domain  encounters
serious difficulties since the integration domains in LFD
amplitudes differ from each other, and it is not easy to restrict
them simultaneously in a self-consistent way. This is a reason why
the PV regularization looks much more preferable. It is naturally
generalized to LFD and allows us to remove  ultra-violet
divergencies in an $\omega$-independent way.  We showed that in
order to cancel completely all $\omega$-dependent terms in the LF
self-energy and the EMV it is necessary to  use the PV
subtractions  for both the boson and fermion propagators. Note
that this cancellation occurs for arbitrary (finite) PV masses.
Simultaneously, all ultra-violet divergencies disappear. After
that, the regularized LFD results coincide exactly with the
Feynman ones calculated with the same PV subtractions.

The physical quantities like the functions ${\cal A}$ and ${\cal
B}$ in the self-energy or the form factors $F_{1,2}$ in the EMV,
can be easily extracted from the corresponding amplitudes. Once we
know the self-energy $\Sigma(p)$ explicitly, taking the traces
\begin{eqnarray*}
\mbox{Tr}\{\Sigma(p)\} & = & 4g^2{\cal A}(p^2),\\
\mbox{Tr} \left\{\Sigma(p){\sla \omega}\right\}&=& {\displaystyle
\frac{4g^2(\omega\cd p)}{m}}{\cal B}(p^2)
\end{eqnarray*}
allows to get ${\cal A}(p^2)$ and ${\cal B}(p^2)$, because the
term $\sim {\sla \omega}$ in the self-energy~(\ref{Sigdecomp})
does not contribute to these traces. To separate the physical
electromagnetic form factors $F_{1,2}$ from the non-physical ones,
$B_{1-3}$, it is enough to consider the plus-component of the
current (or, in terms of covariant LFD, the contraction of the
current with $\omega_{\rho}$) written in the form~(\ref{dec2}).
Indeed,  in the limit $\omega\cd p=\omega\cd p'$,
$$
\bar{u}'\Gamma_{+}u\equiv
\bar{u}'(\omega\cd\Gamma)u=\bar{u}'\left[F_1{\sla
\omega}+\frac{iF_2}{2m}\sigma_{\rho\nu}\omega_{\rho}q_{\nu}\right]u.
$$
As we see, $B_{1-3}$ dropped out from here. This is however not
always the case since the decomposition~(\ref{dec2}) is not
unique. Take, for instance, the decomposition of the same EMV
given in Ref.~\cite{km96}:
\begin{multline}
\label{dec3}
\bar{u}'\Gamma_{\rho}u=\bar{u}'\left\{\tilde{F}_1\gamma_{\rho}+
\frac{i\tilde{F}_2}{2m}\sigma_{\rho\nu}q_{\nu}\right.\\
+B_1\left[\frac{{\sla
\omega}}{\omega\cd
p}-\frac{1}{m(1+\eta)}\right](p+p')_{\rho}\\
\left.+B_2\frac{m\omega_{\rho}}{\omega\cd
p}+B_3\frac{m^2{\sla \omega}\omega_{\rho}}{(\omega\cd
p)^2}\right\}u,
\end{multline}
where $\eta=Q^2/4m^2$. It is easy to see that the form factors
$B_{1-3}$ in Eqs.~(\ref{dec2}) and~(\ref{dec3}) are identical,
while
$$
\tilde{F}_1=F_1-\frac{2\eta
B_1}{1+\eta},\,\,\,\,\,\,\,\,\,\,\tilde{F_2}=F_2-\frac{2B_1}{1+\eta}.
$$
If we now identify $\tilde{F}_{1,2}$ with the physical form
factors, they will differ from $F_{1,2}$, unless $B_1=0$. If we
used for  the regularization the LF cutoffs $\Lambda_{\perp}$ and
$\epsilon$, we would get $B_1=-g^2/64\pi^2$ [see
Eqs.~(\ref{B1noncov}) and~(\ref{B1FLF})] and
\begin{eqnarray*}
\tilde{F}_1 & = &
\frac{g^2}{16\pi^2}\log\frac{\Lambda_{\perp}}{m}+
\frac{g^2}{4\pi^2}\left(\log\frac{m}{\mu}-\frac{7}{8}\right) \\
&&-\frac{g^2Q^2}{24\pi^2m^2}\left(\log\frac{m}{\mu}-\frac{21}{16}\right)
+O(Q^4),\\
\tilde{F}_2 & = &
\frac{7g^2}{32\pi^2}-\frac{5g^2Q^2}{128\pi^2m^2}+O(Q^4).
\end{eqnarray*}
$\tilde{F}_{1,2}$ do not coincide neither with the Feynman form
factors, given by Eqs.~(\ref{F12fey}) (if we identify
$\Lambda_{\perp}$ with $\Lambda$), nor with the LFD ones defined
by Eqs.~(\ref{F1FLF}) and~(\ref{F2FLF}). The  regularization by
the  PV boson kills $B_1$ and this ambiguity disappears. However,
if the non-physical form factors are not canceled completely, the
result is sensitive to the form of the EMV representation.

Although one can use the plus-component of the
current~(\ref{dec2}) to extract the physical form factors in
hadron phenomenology, it is not enough in many other cases, when
the knowledge of the full matrix structure of the vertex is
needed. For example, this occurs in LFD non-perturbative
approaches, when  the  EMV enters as an off-energy-shell subgraph
into a more complicated diagrams. In such a case, one can not
simply ignore the extra spin structures, because they may
contribute to observable quantities. The latter statement concerns
the self-energy as well.

As our results show,  the  dependence on the LF plane orientation
appears not only in LFD amplitudes, but also in Feynman ones, if
we put the  cutoffs in the LF variables ${\bf k}_{\perp}^2$ and
$x$, i.e., constrain the integration domain by a  spherically
non-symmetric  region,  the orientation  of which  follows the
orientation of the LF plane.  At  first glance,  this seems
contradictory: ({\it i}) the extra form factors $B_{1-3}$,
[Eqs.~(\ref{B1FLF})--(\ref{B3FLF})] originate from the spherically
non-symmetric cutoffs, but ({\it ii}) they do not depend on the
cutoff values.  This is in fact a normal situation which can be
illustrated by the following toy example. Consider the
three-dimensional integral
\begin{equation}\label{ex1}
I_{ij}=\int\frac{k_i k_j\,d^3k}{({\bf k}^2+m^2+Q^2)^{5/2}}\equiv
F_1\delta_{ij}.
\end{equation}
It is logarithmically divergent at infinity:  $F_1\sim\int
d{|\bf k}|/|{\bf k}|$.  Let us introduce the spherically
symmetric cutoff $|{\bf k}|<L$. A simple calculation gives
$$
F_1=\frac{2\pi}{3}\log\frac{L^2}{m^2+Q^2}+\frac{4\pi}{9}(\log
8-4)+O\left(\frac{1}{L}\right).
$$
Let us now regularize this integral by a cutoff imposed on the
two-dimensional variable ${\bf k}_{\perp}$ in the plane orthogonal
to an arbitrary direction ${\bf n}$. That is, we  set ${\bf
k}_{\perp}^2={\bf k}^2-({\bf k}\cd {\bf n})^2< \Lambda_{\perp}^2$.
In other words, we integrate over the volume of a cylinder of the
radius $\Lambda_{\perp}$ and of infinite length,  with the axis
directed along ${\bf n}$. The initial integral turns into
\begin{eqnarray}
I^{reg}_{ij}&=&\int\frac{k_i k_j}{({\bf
k}^2+m^2+Q^2)^{5/2}}\theta[\Lambda_{\perp}^2-{\bf k}^2+({\bf k}\cd
{\bf n})^2]\,d^3k \nonumber\\
&\equiv&\tilde{F}_1\delta_{ij}
+B_1(\delta_{ij}-n_i n_j).
\label{ex2}\end{eqnarray}
The integral along ${\bf n}$ (or over $dk_z$, if ${\bf n}$ is
parallel to $z$) converges, similarly to  the  convergence of the
integral over $dx$ in the Feynman amplitude written through the LF
variables. We see that  the spherically non-symmetric cutoff
$\Lambda_{\perp}$ generates one extra form factor $B_1$, like this
happens  for  the Feynman amplitude. We obtain
\begin{eqnarray*}
\tilde{F}_1&=&\frac{2\pi}{3}\log\frac{\Lambda^2_{\perp}}{m^2+Q^2}+
O\left(\frac{1}{\Lambda_{\perp}}\right),
\\
B_1&=&-\frac{2\pi}{3}+O\left(\frac{1}{\Lambda_{\perp}}\right).
\end{eqnarray*}
The leading terms $\propto \log L$ in $F_1$ and $\tilde{F}_1$
coincide, provided we identify $\Lambda_{\perp}$ with $L$.
However, there is a difference in  the  finite parts:
$F_1-\tilde{F}_1=\frac{4\pi}{9}(\log 8-4)$, like the difference
between the term $\frac{15}{16}$ in Eq.~(\ref{F1fey}) and
$\frac{7}{8}$ in Eq.~(\ref{F1FLF}). The transverse cutoff
$\Lambda_{\perp}$ generates  a  finite extra form factor $B_1$,
which itself does not depend on $\Lambda_{\perp}$, similarly to
$B_{1-3}$ in the EMV. This example  clearly mimics  the properties
of the above calculation of the electromagnetic form factors from
the Feynman amplitude regularized by the LF cutoffs.

It is therefore misleading to think that one can derive  some
"true" LF amplitude starting with  the  covariant Feynman
amplitude and calculating it in  the  LF variables: $I=\int \ldots
d^4k=\int\ldots d^2k_{\perp}dx dk_-$. This integral diverges
(except for some particular cases) and it has no sense without
regularization. It depends on the size, shape and orientation of
the integration domain constrained by the regularization
procedure. In other words, there is no "covariant Feynman
amplitude" in itself. There exists only an inseparable couple: the
covariant Feynman amplitude together with the rules to regularize
it.

When we calculate  amplitudes  by means of the LFD graph technique
rules,  the non-invariant integration domain is not the only
source of the $\omega$-dependence. Another source  is hidden in
the rules themselves. Indeed, these rules  are not obliged to
reproduce exactly the amplitude which follows from the Feynman
approach,  even for the same integration domain $D$. The divergent
(and, after  regularization, cutoff-dependent) terms are treated
non-identically  in the two approaches and may be different. This
fact is illustrated by the  above calculation  of the form factor
$B_3$ (cf. Eq.~(\ref{B3noncov}) obtained within LFD with
Eq.~(\ref{B3FLF}) coming from the Feynman approach). The
renormalization procedures  may be different too. However, the
renormalized, observed amplitudes  must  be the same.

When the $\omega$-dependent contributions to the self-energy and
EMV survive, one has to cancel them  with appropriate extra
counterterms in the interaction Hamiltonian. These counterterms
are inherent to the LFD Hamiltonian and should not be confused
with the traditional  charge and mass counterterms in the original
Lagrangian. The need for such a counterterm for the
renormalization of the self-energy was already advocated in
Ref.~\cite{kms_04}, where  the non-perturbative  fermion mass
renormalization was studied in the two-body approximation within
covariant LFD. This new counterterm cancels the contribution $\sim
{\sla \omega}$ in the self-energy~(\ref{Sigdecomp}).

However, the introduction of  additional specific counterterms may
help only if their number is finite and tractable. For the case of
the fermion self-energy we would need only one $\omega$-dependent
counterterm, since the coefficient $\tilde{C}$ is a constant. In
the case of LFD calculations of the EMV,  the form factor $B_3$
depends non-trivially on the momenta (through $Q^2$ and $\alpha$),
which would generate an infinite number of  local  counterterms to
kill it, even in perturbation theory. For this reason, the use of
counterterms can not serve as an universal tool for the
calculation of renormalized quantities in LFD with the traditional
transverse and longitudinal cutoffs.

If  the  renormalization is done correctly, the dependence of any
renormalized amplitude on the cutoffs  disappears and the final
result is the same, regardless to the type of the regularization
used.  However, from the practical point of view, the difficulty
of calculations rapidly increases with introducing extra
counterterms. Therefore, rotationally invariant regularization
(like the PV one) seems by far more preferable  in the LFD
framework. The approach developed in Ref.~\cite{Pierre},
regularizing field theory by defining fields as distributions,
could  provide another method of rotationally invariant
regularization.

\section{\label{concl} Conclusion}
We have calculated perturbatively the fermion self-energy and  the
EMV  in the Yukawa model, in  two different ways:  ({\it i}) by
the LFD  graph  technique rules,  taking into account all
necessary diagrams; ({\it ii}) by  integrating  the Feynman
amplitudes in terms of the LF variables (i.e., the transverse and
longitudinal parts of momenta)  and summing up the pole, arc, and
zero-mode contributions. For the same set of  the  cutoffs imposed
on the LF variables, both methods give different results for the
regularized amplitudes. This difference disappears if the
invariant PV regularization is used instead of the rotationally
non-invariant one.

Such properties follow from the fact that the cutoffs constrain a
spherically non-symmetrical integration domain, the symmetry being
destroyed by the choice of a distinguished direction defined by
the orientation of the LF surface. The dependence of regularized
amplitudes on the LF  plane  orientation is conveniently taken
into account in the explicitly covariant version of LFD by
constructing extra spin structures.  To exclude these structures
from  the  physically observed quantities one may introduce new
counterterms in the interaction Hamiltonian, which also depend on
the LF plane orientation. Taking them into account, one can
calculate  the renormalized amplitudes.

The use of spherically symmetric (in four-dimensional space)
regularization, like, for instance, the PV one, considerably
simplifies calculations. In perturbation theory
it allows to avoid the presence of these extra counterterms at
all.

\begin{acknowledgments}
Two of us (V.A.K. and A.V.S.) are sincerely grateful for the warm
hospitality of the Laboratoire de Physique Corpusculaire,
Universit\'e Blaise Pascal, in Clermont-Ferrand, where the present
study was performed. This work has been partially supported by the
RFBR grant No. 05-02-17482-a.
\end{acknowledgments}
%
\appendix
\section{\label{kin} Kinematical relations}
We give below some kinematical relations used in the self-energy
and EMV calculations within covariant LFD.
Amplitudes of  the LF  diagrams are expressed through
the three-dimensional integrals over $d^3k/\varepsilon_{\bf k}$.
Take the four-vectors $k=(\varepsilon_{\bf k},{\bf k})$ and
$p=(p_0,{\bf p})$, and construct a new four-vector $R=k-xp$ with
$x=(\omega\cd k)/(\omega\cd p)$. Since $\omega\cd R=0$ and
$\omega^2=0$, in the reference frame, where ${\bf p}={\bf 0}$, we
have $\varepsilon_{\bf k}-xp_0-|{\bf k}_{\parallel}|=0$. Taking
into account that $\varepsilon_{\bf k}=\sqrt{{\bf
k}_{\perp}^2+{\bf k}_{\parallel}^2+\mu^2}$ and changing $p_0$ to
$\sqrt{p^2}$, we arrive at the relations
$$
|{\bf k}_{\parallel}|=\frac{{\bf k}_{\perp}^2+\mu^2-x^2p^2}{2x\sqrt{p^2}},\,\,\,\,\,\,\,\,
\varepsilon_{\bf k}=\frac{{\bf k}_{\perp}^2+\mu^2+x^2p^2}{2x\sqrt{p^2}}.
$$
In the variables ${\bf k}_{\perp}$ and $x$, the invariant phase
space element becomes
\begin{equation}
\label{phase} d^3k/\varepsilon_{\bf k}=d^2k_{\perp}dx/x.
\end{equation}

We can now express the scalar product $k\cd p$ entering the
quantity $\tau$, Eq.~(\ref{tau_se}), through ${\bf k}_{\perp}^2$
and $x$. For this purpose, we represent the invariant quantity
$R^2$ in  two different ways:
$$
R^2=(k-xp)^2=\mu^2-2x(k\cd p)+x^2p^2
$$
and
$$
R^2=(\varepsilon_{\bf k}-xp_0)^2-{\bf k}_{\perp}^2-{\bf k}_{\parallel}^2=-{\bf k}_{\perp}^2.
$$
Equating the right-hand sides of these expressions yields
\begin{equation}
\label{kp} k\cd p=\frac{{\bf k}_{\perp}^2+\mu^2+x^2p^2}{2x}.
\end{equation}

In the calculation of the EMV we have one more four-vector $p'$.
We define two new four-vectors $R=k-xp$ and
$R'=k-x'p'$, where $x'=(\omega\cd k)/(\omega\cd p')=x/(1+\alpha)$.
From the equalities $\omega\cd R=\omega\cd R'=0$, in the reference
frame,  where  ${\bf p}={\bf 0}$, follows $R^2=-{\bf
R}_{\perp}^2=-{\bf k}_{\perp}^2$, ${R'}^2=-{{\bf
R}'}_{\perp}^2=-({\bf k}_{\perp}-x'{\bg \Delta})^2$, where ${\bg
\Delta}={\bf p}'_{\perp}={\bf q}_{\perp}$. Analogously to
Eq.~(\ref{kp}), we get from here
\begin{subequations}
\label{kpkpp}
\begin{eqnarray}
k\cd p&=&\frac{{\bf k}_{\perp}^2+\mu^2+x^2m^2}{2x},\\
k\cd p'&=&\frac{({\bf k}_{\perp}-x'{\bg \Delta})^2+\mu^2+{x'}^2m^2}{2x'}.
\end{eqnarray}
\end{subequations}
From these formulas follow the expressions~(\ref{tau})
and~(\ref{taup}). For  completeness, we give here the other scalar
products which are needed  for  the EMV calculations:
\begin{eqnarray}
\omega\cd k&=&x(\omega\cd p),\nonumber \\
\omega\cd p'&=&(1+\alpha)(\omega\cd p),\nonumber \\
p\cd p'&=&\frac{Q^2}{2}+m^2,\nonumber \\
\omega^2&=&0,\,\,\,\,\,\,\,
p^2={p'}^2=m^2,\,\,\,\,\,\,\,k^2=\mu^2.\nonumber
\end{eqnarray}

After integrating over the azimuthal angle of the vector ${\bf
k}_{\perp}$, the form factors depend on ${\bg \Delta}^2$, and one should
relate it with the invariant square of the momentum transfer
$Q^2=-(p'-p)^2$. If one had $\omega\cd p=\omega\cd p'$ (equivalent
to $\alpha=0$), it would be simply $Q^2={\bg \Delta}^2$. In our case the
relation is more complicated. Indeed, since $\omega\cd
p=\omega_0m$, we have, on the one hand, $\omega\cd
p'=(1+\alpha)\omega_0m$, and, on the other hand, $\omega\cd
p'=\omega_0(\sqrt{{\bg \Delta}^2+{{\bf p}'}_{\parallel}^2+m^2}
-|{{\bf p}'}_{\parallel}|)$.
We thus find
$$
|{{\bf p}'}_{\parallel}|=\frac{{\bg \Delta}^2-m^2\alpha(2+\alpha)}{2m(1+\alpha)}
$$
and
\begin{equation}
\label{Qd}
Q^2=(\sqrt{{\bg \Delta}^2+{{\bf p}'}^2_{\parallel}+m^2}-m)^2-
{\bg \Delta}^2-{{\bf p}'}^2_{\parallel}=
\frac{{\bg \Delta}^2+\alpha^2m^2}{1+\alpha}.
\end{equation}
From here Eq.~(\ref{dQ}) follows.
%

\section{\label{sen}Deriving the light-front self-energy from the Feynman approach}
Proceeding from the Feynman amplitude~(\ref{sef}) for the
self-energy, Sec.~\ref{SEF}, we calculate here the pole, arc, and
zero-mode contributions.
 \subsection{Pole and arc contributions}\label{p+arc}
\begin{figure}[ht!]
\begin{center}
\centerline{\includegraphics[scale=0.6]{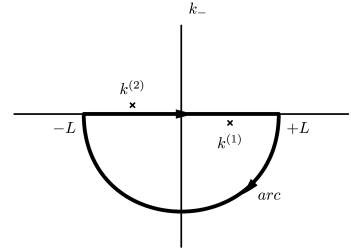}}
\end{center}
 \caption{\label{contour} Contour of the integration over $dk_-$.
Positions of the poles are shown for $\epsilon<x<1-\epsilon$.}
\end{figure}

We first consider the integral over $dk_-$ in Eq.~(\ref{sigma})
for the $x$-integration in the limits $-\infty<x<-\epsilon$,
$\epsilon<x<1-\epsilon$, and $1+\epsilon<x<+\infty$, determining
$\Sigma_{p+a}(p)$. It can be calculated directly (by primitive).
However, in order to get the results in a form closer to LFD, we
will use the residues. The integrand has two poles in the points
\begin{equation*}
k_-=k_-^{(1)}\equiv \frac{{\bf k}_{\perp}^2+\mu^2-i0}{xp_+}\,
\end{equation*}
and
\begin{equation*}
k_-=k_-^{(2)}\equiv p_- -\frac{{\bf k}_{\perp}^2+m^2-i0}
{(1-x)p_+}.
\end{equation*}
If $-\infty<x<-\epsilon$ or $1+\epsilon<x<+\infty$, both poles
lay, respectively, above or below the real axis. If
$\epsilon<x<1-\epsilon$ the poles are situated on the opposite
sides from the real axis. We close the  integration contour by an
arc of a circle of the radius $L$, either in the upper half-plane
or in the lower one, so that the pole $k_-^{(2)}$ is always
outside the contour. The situation is illustrated in
Fig.~\ref{contour}, where the integration contour is shown for the
case $\epsilon<x<1-\epsilon$.

Note that the order of integrations  we have chosen in
Eq.~(\ref{sigma}) requires some care in treating the cutoffs.
Since the integration over $dk_-$ is performed first, we should
keep $L$ arbitrary large, while the other cutoffs $\epsilon$ and
$\Lambda_{\perp}$ are considered as being finite. In other words,
we imply that $L\gg \Lambda^2_{\perp}/(m\epsilon)$. Such a
convention ensures mutual position of the poles and the contour,
as exposed above.
Then $\Sigma_{p+a}(p)$ is represented as
\begin{multline}
\label{SigmaA} \Sigma_{p+a}(p)=\frac{ig^2p_+}{32\pi^4}\int d^2
k_{\perp}\left[\int_{\epsilon}^{1-\epsilon}\left(\Sigma_{pole}-
\Sigma_{arc,\,low}\right)\right. dx\\
- \left. \int_{-\infty}^{-\epsilon}\Sigma_{arc,\,low}dx
-\int_{1+\epsilon}^{+\infty}\Sigma_{arc,\,up}dx\right],
\end{multline}
where $\Sigma_{pole}$ is the residue at the pole $k_-^{(1)}$,
multiplied by $-2\pi i$, while $\Sigma_{arc,\,low}$ and
$\Sigma_{arc,\,up}$ come from the integrations along the arcs in
the lower and upper half-planes, respectively. The result for
$\Sigma_{pole}$ reads
\begin{equation}\label{sig01}
 \Sigma_{pole} =
\frac{2\pi i ({\sla p}-{\sla k}+m)}{p_+[{\bf
k}_{\perp}^2+m^2x-p^2x(1-x)+\mu^2(1-x)]}.
\end{equation}
The four-vector $k$ here is an on-mass-shell four-vector
($k^2=\mu^2$) with the components expressed through ${\bf
k}_{\perp}$ and $x$ as follows: $k\equiv (k_-,k_+,{\bf
k}_{\perp})= (\frac{{\bf k}_{\perp}^2+\mu^2}{xp_+},xp_+,{\bf
k}_{\perp})$. To calculate $\Sigma_{arc,\,low}$ and
$\Sigma_{arc,\,up}$, we should move along the arc in the clockwise
and counter-clockwise directions, respectively. In the points of
the arc $k_-=Le^{i\phi}$ and $dk_-=iLe^{i\phi}d\phi$, where $\phi$
is an azimuthal angle. In the limit $L\to\infty$, we retain in the
integrand the dominating $k_-$-term. We thus obtain
\begin{equation}
\label{Sarclow} \Sigma_{arc,\,low}=-\Sigma_{arc,\,up}=-\frac{\pi
i}{2p_+^2}\frac{\gamma_+}{x(1-x)}.
\end{equation}
Substituting Eqs.~(\ref{sig01}) and~(\ref{Sarclow}) into
Eq.~(\ref{SigmaA}) and going over to the explicitly covariant
notations by means of the identities $\gamma_+={\sla \omega}$,
$p_+=\omega\cd p$, we obtain that $\Sigma_{p+a}(p)$ is given by
Eq.~(\ref{SigmaA1}).
\subsection{Zero modes}\label{zmc}
Consider now the zero-mode term $\Sigma_{zm}(p)$.  We denote
$$
\Sigma_{zm}(p)\equiv \Sigma_{zm}^{(0)}(p)+\Sigma_{zm}^{(1)}(p),
$$
where the two items on the right-hand side correspond to the
contributions  to Eq.~(\ref{sigma})  from the infinitesimal
integration  regions $-\epsilon<x<\epsilon$ and
$1-\epsilon<x<1+\epsilon$, respectively. Take first
$\Sigma_{zm}^{(0)}(p)$ (bosonic zero modes). If $x=0$, the term
$k_-p_+x$ in the denominator of the integrand disappears and the
integral over $dk_-$ diverges linearly. This infinite contribution
should therefore be proportional to $\delta(x)$. We take $x\to 0$
and keep  in the numerator and in the denominator of the integrand
in Eq.~(\ref{sigma})  the leading $k_-$-terms (where $k_-$ is not
multiplied by $x$) only. That is
\begin{multline}
\label{sigma4}
 \Sigma_{zm}^{(0)} = \frac{ig^2p_+}{32\pi^4}\int d^2k_{\perp}
 \int_{-\epsilon}^{\epsilon} dx \\
 \times\int_{-L}^L dk_-
 \frac{1}{[k_- p_+ x
-{\bf k}_{\perp}^2-\mu^2+i0]}
 \frac{-\frac{1}{2}\gamma_+k_-}{(-k_-p_+)}.
\end{multline}
The double integral over $dxdk_-$ is of the type of
Eq.~(\ref{apb1}) with $b=({\bf k}_{\perp}^2+\mu^2)/p_+$ and $h=0$
(see Appendix~\ref{appB}). Using Eq.~(\ref{intzm1}), we obtain, in
the covariant notations:
\begin{multline}
\Sigma_{zm}^{(0)}(p) =  \frac{g^2{\sla
\omega}}{32\pi^3(\omega\cd p)} \int d^2 k_{\perp}\\
\times\left[\log\frac{(\omega\cd p)L }{{\bf k}_{\perp}^2+\mu^2}
-\log\frac{1}{\epsilon}-\frac{i\pi}{2}\right]. \label{sigma4a}
\end{multline}
Changing the variable $x\to 1-x$, for $\Sigma_{zm}^{(1)}(p)$
(fermionic zero modes) we similarly get
\begin{multline}
\Sigma_{zm}^{(1)}(p)= - \frac{g^2{\sla
\omega}}{32\pi^3(\omega\cd p)} \int d^2 k_{\perp}\\
\times\left[\log\frac{(\omega\cd p)L }{{\bf k}_{\perp}^2+m^2}
-\log\frac{1}{\epsilon}-\frac{i\pi}{2}\right]. \label{sigma5}
\end{multline}
Taking the sum of (\ref{sigma4a}) and (\ref{sigma5}), we find that
it has the form of eq. (\ref{sigma0}): $\Sigma_{zm}(p)\equiv
g^2C_{zm}\frac{m{\sla \omega}}{\omega\cd p}$ with
$$
C_{zm}=\frac{g^2}{32\pi^2}  \int^{\Lambda_{\perp}^2}_0 d{\bf
k}_{\perp}^2
 \; \log\frac{{\bf k}_{\perp}^2+m^2} {{\bf k}_{\perp}^2+\mu^2}
$$
Calculating this integral, we find Eq.~(\ref{CZM}) for $C_{zm}$.

\section{\label{ffs} Light-front contributions to the form factors}

In this Appendix, starting with the LF vertices
$\Gamma_{\rho}^{(A)}$, $\Gamma_{\rho}^{(B)}$, and
$\Gamma_{\rho}^{(C)}$ found in Sec.~\ref{LFvert},
Eqs.~(\ref{GmA})--(\ref{GmC}), we calculate their contributions to
form factors $F_{1,2}$, $B_{1-3}$.

\subsection{\label{prtr}Preliminary transformations}

Each contribution to a form factor ${\cal F}$ (${\cal F}\equiv
F_i$ or $B_i$) from the vertices $\Gamma_{\rho}^{(A)}$,
$\Gamma_{\rho}^{(B)}$, or $\Gamma_{\rho}^{(C)}$ can be written as
an integral of the form
\begin{equation}
\label{ffint3d} {\cal F}=\int_{x_{min}}^{x_{max}}
dx\int_0^{\Lambda_{\perp}^2} d{\bf
k}_{\perp}^2\int_0^{2\pi}d\phi\,f({\bf
k}^2_{\perp},x,\phi,Q^2,\alpha),
\end{equation}
where $\phi$ is the angle between the two-dimensional vectors
${\bf k}_{\perp}$ and ${\bg \Delta}$, $f$ is  a  given function
($f$'s are different for different form factors), and the modulus
of ${\bg \Delta}$ is expressed through $Q^2$ by Eq.~(\ref{dQ}).
The limits $x_{min}$ and $x_{max}$ depend on which vertex is
considered.

The  analysis of the dependence of the form factors on the cutoff
$\Lambda_{\perp}$ (when it tends to infinity) allows to represent
each of them as
\begin{equation}
\label{FFc} {\cal
F}=a_{1}\Lambda_{\perp}^2+a_{2}\log\frac{\Lambda_{\perp}}{m}+a_{reg},
\end{equation}
where the coefficients $a_{1}$, $a_{2}$, and $a_{reg}$ are regular
functions depending on  the particle masses, $Q^2$, and $\alpha$,
but independent of $\Lambda_{\perp}$. If $\alpha\neq 0$, no other
divergencies (i.e., cutoff-dependent terms) excepting those listed
in Eq.~(\ref{FFc}) appear. Since the coefficients $a_{1}$ and
$a_{2}$ determine  the dependence of the form factors on the
cutoff, we will calculate them  at arbitrary $\alpha$, in order to
see in detail how the cutoff disappears (if so!) after taking into
account all the contributions $\Gamma_{\rho}^{(A)}$,
$\Gamma_{\rho}^{(B)}$, and $\Gamma_{\rho}^{(C)}$ to the full EMV
and its subsequent renormalization. As far as the coefficients
$a_{reg}$ are concerned, we will  find them in the limit
$\alpha\to 0$, retaining  non-vanishing terms only. This
simplifies calculations a lot, but  leads to  the same final
result as  if $\alpha$ was non-zero.

Although direct integrations over $d\phi$ and $d{\bf k}_{\perp}^2$
can always be done analytically for arbitrary $Q^2$ and $\alpha$,
the results turn out to be very cumbersome, so that the final
integration over $dx$ (if it can be done analytically) takes too
much time. To make  the  computations more effective, we will
proceed in the following way. We take  the initial expression for
a  given form factor in the form of Eq.~(\ref{ffint3d}) and study
the behavior of the integrand as a function of ${\bf k}^2_{\perp}$
in the asymptotic region $|{\bf k}_{\perp}|\to \infty$. For this
purpose we decompose it in a Laurent series:
\begin{multline}
\label{ftLs} f({\bf k}^2_{\perp},x,\phi,{Q^2},\alpha)=
f_0(x,Q^2,\alpha)\\
+\frac{f_1(x,{Q^2},\alpha)\cos\phi}{|{\bf k}_{\perp}|}+
\frac{f_2(x,\phi,{Q^2},\alpha)}{{\bf k}_{\perp}^2}+\ldots.
\end{multline}
We use this decomposition in order to define the functions $f_0$
and $f_2$ (the term with $f_1$ drops out after the integration
over $d\phi$). We then represent $f$ as a sum
\begin{multline}
\label{ft1} f({\bf k}^2_{\perp},x,\phi,{Q^2},\alpha)\equiv
f_{reg}({\bf k}^2_{\perp},x,\phi,Q^2,\alpha)\\
+f_0(x,Q^2,\alpha)+\frac{\Phi_2(x,Q^2,\alpha)} {{\bf
k}_{\perp}^2+m^2\beta(x)},
\end{multline}
where
\begin{equation}
\label{Phi} \Phi_2(x,Q^2,\alpha)= \frac{1}{2\pi}\int_0^{2\pi}
f_2(x,\phi,Q^2,\alpha)d\phi
\end{equation}
and  $\beta(x)$ is a positive function which can be chosen in any
convenient way in order to avoid the singularity at ${\bf
k}_{\perp}^2\to 0$. Eq.~(\ref{ft1}) should be considered as a
definition of the function $f_{reg}({\bf
k}_{\perp}^2,x,\phi,Q^2,\alpha)$. The latter, after the
integration over $d\phi$, decreases faster than $1/{\bf
k}_{\perp}^2$ in the asymptotic region, and its integration over
$d{\bf k}_{\perp}^2$ does not require  any regularization.  The
functions $f_0(x,Q^2,\alpha)$ and $\Phi_2(x,Q^2,\alpha)$ are
rather simple, so that the two last addenda on the right-hand side
of Eq.~(\ref{ft1}) can be easily integrated over all the
variables.

After separating  out the terms which slowly decrease  when $|{\bf
k}_{\perp}|\to \infty$, we remain with a set of regular functions
$f_{reg}$. At arbitrary $\alpha$, these functions are even more
complicated than the initial integrands $f$. However, we calculate
the regular contributions to the form factors in the limit
$\alpha\to 0$. So, it is enough to consider the limiting functions
$f_{reg}({\bf k}^2_{\perp},x,\phi,Q^2,\alpha\to 0)$ which are much
simpler than $f({\bf k}^2_{\perp},x,\phi,Q^2,\alpha)$. If
$f_{reg}$ is singular at $\alpha\to 0$, we decompose it in powers
of $\alpha$ and retain all non-vanishing terms.

The trick exposed above allows to  find  the coefficients in
Eq.~(\ref{FFc}). Indeed, substituting Eq.~(\ref{ft1}) into
Eq.~(\ref{ffint3d}), we get
\begin{subequations}
\label{L1reg}
\begin{eqnarray}
a_{1} & = & 2\pi\int_{x_{min}}^{x_{max}}f_0(x,Q^2,\alpha) dx,
\label{L1} \\
a_{2} & = &
4\pi\int_{x_{min}}^{x_{max}}\Phi_2(x,Q^2,\alpha)dx, \label{L2}\\
a_{reg} & = & \int_{x_{min}}^{x_{max}}dx\,\left\{\left[
\int_0^{\infty}d{{\bf
k}_{\perp}^2}\right.\right. \nonumber \\
&&\times\left.\int_0^{2\pi}d\phi\, f_{reg}({\bf
k}_{\perp}^2,x,\phi,Q^2,\alpha\to
0)\right]\nonumber\\
&& \left. \phantom{\int}-2\pi\Phi_2(x,Q^2,\alpha\to
0)\log\beta(x)\right\}.\label{reg}
\end{eqnarray}
\end{subequations}
%
The integrations over $d\phi$ and $d{\bf k}_{\perp}^2$ can be done
analytically for all the five form factors. For $F_{1,2}$ the
results of  the remaining integrations over $dx$ at arbitrary
$Q^2$ and $\mu$ are not expressed through elementary functions. In
order to simplify the formulas, we decompose both $F_{1,2}$ in
powers of $Q^2$ up to terms of order $Q^2$ and take the limit
$\mu\to 0$. Concerning the form factors $B_{1-3}$, all  the
integrations are performed analytically without any approximation.

Below, in Secs.~\ref{GA}--\ref{GC}, we list  the coefficients
$a_{1}$, $a_{2}$, and $a_{reg}$ entering Eq.~(\ref{FFc}), pointing
out in superscripts which form factor the given coefficient
belongs to. The coefficients are calculated by means of
Eqs.~(\ref{L1reg}), separately for the vertices
$\Gamma_{\rho}^{(A)}$, $\Gamma_{\rho}^{(B)}$, and
$\Gamma_{\rho}^{(C)}$ [see Eqs.~(\ref{GLFT}) and~(\ref{GLFC})].
\subsection{\label{GA} Contribution to the form factors from
$\Gamma_{\rho}^{(A)}$}
It is convenient to  set  in Eq.~(\ref{ft1})
$\beta(x)=x^2+\frac{\mu^2}{m^2}(1-x)$. The limits of the
$x$-integration are $x_{min}=0$, $x_{max}=1$. Then
\begin{equation}
\label{aL1A}a_{1}^{F_1}=a_{1}^{F_2}=a_{1}^{B_1}=a_{1}^{B_2}=a_{1}^{B_3}=0,
\end{equation}
\begin{widetext}
\begin{eqnarray}
a_{2}^{F_1} & = & {\displaystyle
\frac{g^2\left[2(1+\alpha)Q^2+\alpha^2(2-\alpha^2)m^2
\right]}{16\pi^2(1+\alpha)^2z_{\alpha}}},\quad
a_{2}^{F_2} = 0,
\nonumber \\
a_{2}^{B_1} &=& {\displaystyle
\frac{g^2(2+\alpha)\left[(1+\alpha)Q^2+\alpha^2m^2
\right]}{16\pi^2(1+\alpha)^2z_{\alpha}},}
\label{aL2A} \\
a_{2}^{B_2} & = & {\displaystyle \frac{g^2}{16\pi^2(1+\alpha)}
\left\{\frac{4\alpha^2(2+2\alpha+\alpha^2)m^2+[8+\alpha(2+\alpha)(8+2\alpha+\alpha^2)]Q^2}
{(1+\alpha)(2+\alpha)z_{\alpha}}-4\right\},}\nonumber \\
a_{2}^{B_3} & = & {\displaystyle
\frac{g^2}{16\pi^2m^2(1+\alpha)^2}
\left\{(2+\alpha)\mu^2-\frac{(3+3\alpha+\alpha^2)m^2}{1+\alpha}
-\frac{Q^2[(1+\alpha)Q^2-\alpha^2m^2]}{z_{\alpha}}\right\},}\nonumber
\end{eqnarray}
 where $z_{\alpha}=(2+2\alpha+\alpha^2)Q^2+2\alpha^2m^2$, and
\begin{eqnarray}
a_{reg}^{F_1} & = &
\frac{g^2}{4\pi^2}\left(\log\frac{m}{\mu}-\frac{7}{8}\right)-\frac{g^2Q^2}{24\pi^2m^2}\left(\log\frac{m}{\mu}-\frac{9}{8}\right)+O(Q^4),
\nonumber \\
 a_{reg}^{F_2} & = &
\frac{3g^2}{16\pi^2}-\frac{g^2Q^2}{32\pi^2m^2}+O(Q^4),
\nonumber \\
a_{reg}^{B_1} & = &
\frac{g^2}{64\pi^2}\left[-1+4\varphi(Q^2)\right],
\label{aregA}\\
a_{reg}^{B_2} & = &
-\frac{g^2}{32\pi^2}\left[1+4\varphi(Q^2)\right],\nonumber \\
a_{reg}^{B_3} & = &
\frac{g^2}{32\pi^2m^2}\left[\mu^2\left(2\log\frac{m}{\mu}+1\right)\phantom{\frac{1}{1}}-(4m^2-2\mu^2+Q^2)\varphi(Q^2)\right],\nonumber
\end{eqnarray}
\end{widetext}
with
\begin{equation}
\label{funphi} \varphi(Q^2)=1-\sqrt{1+\frac{4m^2}{Q^2}}
\log\left(\frac{\sqrt{Q^2+4m^2}+\sqrt{Q^2}}{2m}\right).
\end{equation}
\subsection{\label{GB} Contribution to the form factors from
$\Gamma_{\rho}^{(B)}$}
We set $\beta(x)=1$, $x_{min}=1$, $x_{max}=1+\alpha$. We thus get
\begin{widetext}
\begin{eqnarray}
a_{1}^{F_1}&=&a_{1}^{F_2}=a_{1}^{B_1}=a_{1}^{B_2}=0,\nonumber \\
a_{1}^{B_3}&=&-\frac{g^2}{32\pi^2m^2}\,\frac{\log(1+\alpha)}{\alpha(1+\alpha)},\nonumber\\
a_{2}^{F_1} & = & {\displaystyle
\frac{g^2}{16\pi^2}\left[1-\frac{2(1+\alpha)Q^2+\alpha^2(2-\alpha^2)m^2}
{(1+\alpha)^2z_{\alpha}}\right]},\quad
a_{2}^{F_2} = 0,\label{aL1B}\\
a_{2}^{B_1} & = & {\displaystyle
-\frac{g^2(2+\alpha)\left[(1+\alpha)Q^2+\alpha^2m^2
\right]}{16\pi^2(1+\alpha)^2z_{\alpha}},}\nonumber \\
a_{2}^{B_2} & = & {\displaystyle -\frac{g^2}{16\pi^2(1+\alpha)}
\left\{\frac{4\alpha^2(2+2\alpha+\alpha^2)m^2+[8+\alpha(2+\alpha)(8+2\alpha+\alpha^2)]Q^2}
{(1+\alpha)(2+\alpha)z_{\alpha}}-4\right\},}\nonumber\\
a_{2}^{B_3} & = & {\displaystyle
-\frac{g^2}{16\pi^2m^2(1+\alpha)^2}
\left\{\mu^2-\frac{(2+\alpha)m^2}{1+\alpha}
-\frac{Q^2[(1+\alpha)Q^2-\alpha^2m^2]}{z_{\alpha}}\right\}.}\nonumber
\end{eqnarray}
\end{widetext}

While integrating over $dx$ in Eq.~(\ref{reg}), we introduce a new variable
$\xi=(x-1)/\alpha$. Then
$$
\int_{1}^{1+\alpha}dx(\ldots)=\alpha\int_0^1 d\xi(\ldots).
$$
After that the integrands $f_{reg}$ should be decomposed in powers
of $\alpha$ up to terms of order $1/\alpha$. Performing
this transformation, we obtain the following result:
\begin{eqnarray}
a_{reg}^{F_1} & = &  a_{reg}^{F_2} = 0,\nonumber \\
a_{reg}^{B_1} & = & {\displaystyle -\frac{1}{2}a_{reg}^{B_2} =
-\frac{g^2}{16\pi^2}\varphi(Q^2),} \\
a_{reg}^{B_3} & = & {\displaystyle
\frac{g^2}{32\pi^2m^2}\left[\frac{Q^2}{\alpha}+(4m^2-2\mu^2+Q^2)\varphi(Q^2)\right]}\nonumber
\end{eqnarray}
with $\varphi(Q^2)$ given by Eq.~(\ref{funphi}).
%
\subsection{\label{GC}Contribution to the form factors from
$\Gamma_{\rho}^{(C)}$}
The matrix structure of the vertex $\Gamma_{\rho}^{(C)}$ is the
same as the one in front of the form factor $B_3$ in the
decomposition~(\ref{dec2}). For this reason $\Gamma_{\rho}^{(C)}$
contributes to $B_3$ only. From Eq.~(\ref{GLFC1}) we easily find
\begin{equation}
\label{aL1C}
a_{1}^{B_3}=\frac{g^2}{32\pi^2m^2}\,\frac{\log(1+\alpha)}{\alpha(1+\alpha)},
\end{equation}
all the other coefficients being zero.
%

\section{\label{derEMV}Deriving the light-front electromagnetic vertex
from the Feynman approach} Proceeding from the Feynman
amplitude~(\ref{eq1}) for the EMV, Sec.~\ref{EMFey}, we calculate
here the pole, arc, and zero-mode contributions.
\subsection{\label{prelim}Preliminary transformations}
Under the condition $p'_+-p_+=0$, the LF components of the
four-vectors $p$ and $p'$ are
\begin{equation}
\label{ppp}
\begin{array}{ll}
 {\bf p}_{\perp}=0, & p_-={\displaystyle \frac{m^2}{p_+}},\\
 {\bf p}'_{\perp}={\bf q}_{\perp}, & p'_-={\displaystyle
 \frac{{\bf q}_{\perp}^2+m^2}{p_+}},
\end{array}
\end{equation}
where ${\bf q}_{\perp}$ is the transversal part of the
three-dimensional momentum transfer:
$$
\label{qqq} {\bf q}_{\perp}={\bf p}'_{\perp}-{\bf p}_{\perp},\quad
Q^2=-(p'-p)^2={\bf q}_{\perp}^2.
$$
Introducing a new variable $x$ by means of the relation $k_+=xp_+$
and using the notations
\begin{equation}\label{abbp}
 a=\frac{{\bf k}_{\perp}^2+\mu^2}{p_+},\quad
b=\frac{{\bf k}_{\perp}^2+m^2}{p_+},\quad b'=\frac{({\bf
k}_{\perp}-{\bf q}_{\perp})^2+m^2}{p_+},
\end{equation}
we cast Eq.~(\ref{eq2}) in the form
\begin{multline}
\label{eq3} \Gamma_{\rho}=\frac{ig^2}{128\pi^4 p_+^2}\int
d^2k_{\perp} \int_{-\infty}^{+\infty}dx\\
\times\int_{-\infty}^{+\infty} \frac{dk_-\,[k_-\gamma_++ {\cal
M}']\gamma_{\rho}[k_-\gamma_+ +{\cal M}]} {v_1v_2v'_2},
\end{multline}
where
\begin{eqnarray}
v_1&=&xk_--a+i0,\nonumber \\
v_2&=&(x-1)k_--b-(x-1)p_-+i0,\nonumber \\
v_2'&=&(x-1)k_--b'-(x-1)p'_-+i0,\label{eq4}\\
\nonumber \\
 {\cal M}&=&(x-1)p_+\gamma_--p_-\gamma_+-2{\bf
k}_{\perp}\cd{\bg \gamma}_{\perp}-2m,\label{eq5} \\
{\cal M}'&=&(x-1)p_+\gamma_--p'_-\gamma_+-2({\bf k}_{\perp}-{\bf
q}_{\perp})\cd{\bg \gamma}_{\perp}-2m. \nonumber
\end{eqnarray}
The matrices ${\cal M}$ and ${\cal M}'$ do not depend on $k_-$.

The integral in Eq.~(\ref{eq3}) diverges at $|{\bf
k}_{\perp}|\to\infty$. If $x=0$ or $x=1$, it diverges also at
$k_-\to\pm\infty$. We constrain the integration over $d{\bf
k}_{\perp}^2$ in $d^2k_{\perp}=\frac{1}{2}d{\bf k}_{\perp}^2d\phi$
by the cutoff $\Lambda_{\perp}^2$, and introduce also the cutoffs
$L$ in $k_-$, as in Eq.~(\ref{Lkm}), and $\epsilon$ in $x$,
splitting the region of the integration over $dx$ into three
"normal" regions $-\infty<x<-\epsilon$, $\epsilon<x<1-\epsilon$,
$1+\epsilon<x<+\infty$ and two infinitesimal ones
$-\epsilon<x<\epsilon$, $1-\epsilon<x<1+\epsilon$. The integrals
over  the "normal" regions are calculated by summing up the
residue and arc contributions, as in Sec.~\ref{Sigma_Fey}. The
contributions (if any) from the two infinitesimal regions of the
integration over $dx$, which do not vanish in the limit
$\epsilon\to 0$, correspond to the zero modes. Because of rather
weak (logarithmic) divergence of the initial integral~(\ref{eq1})
we expect that the result of the full integration over $dk_-$ and
$dx$ has a finite limit at $L\to\infty$ and $\epsilon\to 0$, while
the remaining integration over $d^2k_{\perp}$ produces logarithmic
dependence of the EMV on the cutoff $\Lambda_{\perp}$.

%
 \subsection{Pole and arc contributions}\label{parcEMV}
We start with the calculation of $\Gamma_{\rho}^{(p+a)}$. The
procedure is quite similar to that exposed in Appendix~\ref{p+arc}
and in the papers~\cite{lb95, Nico1, jaus_2003, bakker_05,
jibak1}. The integrand in Eq.~(\ref{eq3}) has three poles:
\begin{eqnarray}
k_-^{(1)}&=&\frac{a-i0}{x},\nonumber \\
k_-^{(2)}&=&\frac{b+(x-1)p_--i0}{x-1},\nonumber \\
k_-^{(3)}&=&\frac{b'+(x-1)p'_--i0}{x-1}.\label{poles}
\end{eqnarray}
If $-\infty<x<-\epsilon$ or $1+\epsilon<x<+\infty$, all  of them
are, respectively, either in the upper half-plane of $k_-$ or in
the lower one. If $\epsilon<x<1-\epsilon$, the pole $k_-^{(1)}$ is
in the lower half-plane, while the poles $k_-^{(2)}$ and
$k_-^{(3)}$ are in the upper one. We can thus write, analogously
to Eq.~(\ref{SigmaA}):
\begin{multline}
\label{intkm} \Gamma_{\rho}^{(p+a)}=\frac{ig^2}{128\pi^4p_+^2}
\int d^2k_{\perp}\left[\int_{\epsilon}^{1-\epsilon}\left(
\Gamma_{\rho}^{pole}-\Gamma_{\rho}^{arc,\,low}\right)dx\right.\\
\left.-\int_{-\infty}
^{-\epsilon}\Gamma_{\rho}^{arc,\,low}dx-\int_{1+\epsilon}
^{+\infty}\Gamma_{\rho}^{arc,\,up}dx\right],
\end{multline}
where $\Gamma_{\rho}^{pole}$ equals the residue of the integrand
in the pole $k_-=k_-^{(1)}$, multiplied by $-2\pi i$:
\begin{equation}
\label{residue} \Gamma_{\rho}^{pole}=-\frac{8\pi
ip_+^2}{x}\,\frac{({\sla p}'-{\sla k}+m)\gamma_{\rho}({\sla
p}-{\sla k}+m)}{[(p'-k)^2-m^2][(p-k)^2-m^2]}
\end{equation}
with $k^2=\mu^2$, and the arc contributions are
\begin{equation}
\label{IupIlow}
\Gamma_{\rho}^{arc,\,low}=-\Gamma_{\rho}^{arc,\,up}=-\frac{\pi
i}{x(1-x)^2}\,\gamma_+\gamma_{\rho}\gamma_+.
\end{equation}
Substituting Eqs.~(\ref{residue}) and~(\ref{IupIlow}) into
Eq.~(\ref{intkm}), we  find
\begin{widetext}
\begin{eqnarray}
\Gamma_{\rho}^{(p+a)}&=&\frac{g^2}{16\pi^3}\int
d^2k_{\perp}\left\{\int_{\epsilon}^{1-\epsilon}\frac{dx}{x}\,
\frac{({\sla p}'-{\sla k}+m)\gamma_{\rho}({\sla p}-{\sla
k}+m)}{[(p'-k)^2-m^2][(p-k)^2-m^2]}-\frac{\gamma_+\gamma_{\rho}\gamma_+}{4p_+^2}
\log\frac{1}{\epsilon} \right\} \label{GLF0}
\end{eqnarray}
\end{widetext}
We substitute here $\log(1/\epsilon)$ by $\int_{\epsilon}^{1}dx/x$
and take the limit $\epsilon\to 0$. The latter is achieved  simply
by setting $\epsilon=0$ in the integration limits, because the
integrand is no more singular neither at $x=0$ nor at $x=1$. We
also return to the covariant notations, replacing $\gamma_+$ by
$\sla{\omega}$ and $p_+$ by $\omega\cd p$. In this way we find
that $\Gamma_{\rho}^{(p+a)}$ is given by Eq.~(\ref{GLF1}).


\subsection{\label{zmEMV} Zero modes}

Let us now consider  the zero mode contribution to
Eq.~(\ref{eq3}). It can be written as
\begin{equation}
\label{eq4a} \Gamma_{\rho}^{(zm)}=\frac{ig^2}{128\pi^4p_+^2}\int
d^2k_{\perp}\left[{\cal G}_{\rho}^{(x=0)}+{\cal
G}_{\rho}^{(x=1)}\right],
\end{equation}
where
\begin{subequations}
\label{eqzm01}
\begin{eqnarray}
\label{eqzm0} {\cal
G}_{\rho}^{(x=0)}&=&\int_{-\epsilon}^{\epsilon}dx\int_{-L}^{L}
\frac{dk_-\,[k_-\gamma_++ {\cal
M}']\gamma_{\rho}[k_-\gamma_++{\cal M}]} {v_1v_2v'_2},
\nonumber \\
\\
{\cal
G}_{\rho}^{(x=1)}&=&\int_{1-\epsilon}^{1+\epsilon}dx\int_{-L}^{L}
\frac{dk_-\,[k_-\gamma_++ {\cal
M}']\gamma_{\rho}[k_-\gamma_++{\cal M}]} {v_1v_2v'_2}.\nonumber \\
\label{eqzm1}
\end{eqnarray}
\end{subequations}
In order to calculate these integrals, we need their asymptotical
limit when $L\to\infty$ and $\epsilon\to 0$. As explained in
Sec.~\ref{Sigma_Fey}, one should take the limit $L\to\infty$
first, while keeping $\epsilon$ finite, and then allow
$\epsilon\to 0$.

We represent ${\cal G}_{\rho}^{(x=0)}$ as
\begin{multline*}
{\cal G}_{\rho}^{(x=0)}=
\int_{-\epsilon}^{\epsilon}dx\int_{-L}^{L}dk_-\,\left\{
\gamma_+\gamma_{\rho}\gamma_+\,\frac{k_-^2} {v_1v_2v'_2}\right.\\
+ ({\cal
M}'\gamma_{\rho}\gamma_++\gamma_+\gamma_{\rho}{\cal
M})\,\frac{k_-}{v_1v_2v'_2}
\left.+ {\cal M}'\gamma_{\rho}{\cal
M}\,\frac{1}{v_1v_2v'_2}\right\}.
\end{multline*}
For further calculations, we will make use of the formulas
\begin{equation}\label{eq6}
 \frac{k_-^n}{v_1v_2v'_2}=\frac{D_n(x)}{v_1}
+\frac{E_n(x)}{v_2}+\frac{F_n(x)}{v'_2},\quad n=0,\,1,\,2,
\end{equation}
where $v_1$, $v_2$, and $v'_2$ are defined by Eqs.~(\ref{eq4}),
and the functions $D_n(x)$, $E_n(x)$, and $F_n(x)$ are
\begin{widetext}
\begin{eqnarray}
 D_0(x)={\displaystyle \frac{x^2}{t_1t_1'}},\quad
 &E_0(x)={\displaystyle \frac{x-1}{t_1t_2}},\quad
 &F_0(x)={\displaystyle -\frac{x-1}{t_1't_2}}, \nonumber\\
 D_1(x)={\displaystyle \frac{ax}{t_1t_1'}},\quad
 &E_1(x)={\displaystyle \frac{b+(x-1)p_-}{t_1t_2}},\quad
 &F_1(x)={\displaystyle -\frac{b'+(x-1)p'_-}{t_1't_2}}, \label{apa1} \\
 D_2(x)={\displaystyle \frac{a^2}{t_1t_1'}},\quad
 &E_2(x)={\displaystyle \frac{[b+(x-1)p_-]^2}{(x-1)t_1t_2}},\quad
 &F_2(x)={\displaystyle -\frac{[b'+(x-1)p'_-]^2}{(x-1)t_1't_2}},\nonumber
\end{eqnarray}
\end{widetext}
with
\begin{eqnarray*}
t_1&=&(x-1)(xp_--a)+bx,\\
 t'_1&=&(x-1)(xp'_--a)+b'x,\\
t_2&=&b-b'+(p_--p'_-)(x-1).
\end{eqnarray*}
The values of $a$, $b$, and $b'$ are defined by Eqs.~(\ref{abbp}).

After the
transformation~(\ref{eq6}), the problem of finding ${\cal
G}_{\rho}^{(x=0)}$ reduces to the calculation of integrals of the
following three types:
\begin{eqnarray}
&& \int_{-\epsilon}^{\epsilon}dx\,f(x)\int_{-L}^{L}
\frac{dk_-} {v_1},\nonumber \\
&&\int_{-\epsilon}^{\epsilon}dx\,f(x)\int_{-L}^{L} \frac{dk_-}
{v_2},\label{3t} \\
&&\int_{-\epsilon}^{\epsilon}dx\,f(x)\int_{-L}^{L}
\frac{dk_-} {v'_2},\nonumber
\end{eqnarray}
with various functions $f(x)$, being either $D_n(x)$, $E_n(x)$,
and $F_n(x)$, or their products with ${\cal M}$ and ${\cal M}'$.
It is easy to see that the latter two integrals in Eqs.~(\ref{3t})
always give zero in the limit $\epsilon\to 0$, while the first
one is of the type of the integral~(\ref{apb3}) from
Appendix~\ref{appB}. Using the formula~(\ref{apb3}) and taking
into account that $D_2(0)=1$, we get
\begin{eqnarray}
 {\cal G}_{\rho}^{(x=0)}&=&\gamma_+\gamma_{\rho}\gamma_+
\int_{-\epsilon}^{\epsilon}dx\,D_2(x)\int_{-L}^{L}
\frac{dk_-}{v_1} \nonumber \\
&=&-\gamma_+\gamma_{\rho}\gamma_+\left[\pi^2+2\pi
i\left(\log\frac{L}{a}-\log\frac{1}{\epsilon}\right)\right].\nonumber \\
\label{eq7}
\end{eqnarray}
 The  calculation of ${\cal G}_{\rho}^{(x=1)}$ is
similar, but  the algebra is  more lengthy. Changing the
variable $x\to x-1$, we have
\begin{multline*}
 {\cal G}_{\rho}^{(x=1)}=
\int_{-\epsilon}^{\epsilon}dx\int_{-L}^{L}dk_-\,\left\{
\gamma_+\gamma_{\rho}\gamma_+\,\frac{k_-^2} {u_1u_2u'_2}\right.\\
+ ({\cal M}_1'\gamma_{\rho}\gamma_++\gamma_+\gamma_{\rho}{\cal
M}_1)\,\frac{k_-}{u_1u_2u'_2}\\
\left.+ {\cal M}_1'\gamma_{\rho}{\cal
M}_1\,\frac{1}{u_1u_2u'_2}\right\},
\end{multline*}
where
\begin{eqnarray}
u_1&=&(x+1)k_--a+i0,\nonumber \\
u_2&=&xk_--b-xp_-+i0,\nonumber \\
u_2'&=&xk_--b'-xp'_-+i0,\label{eq11}\\
\nonumber\\
{\cal M}_1&=&xp_+\gamma_--p_-\gamma_+-2{\bf
k}_{\perp}\cd {\bg \gamma}_{\perp}-2m,\label{eq12}\\
{\cal M}'_1&=&xp_+\gamma_--p'_-\gamma_+-2({\bf k}_{\perp}-{\bf
q}_{\perp})\cd {\bg \gamma}_{\perp}-2m. \nonumber
\end{eqnarray}
We have now, instead of Eq.~(\ref{eq6}):
\begin{eqnarray}
\label{eq13} \frac{k_-^n}{u_1u_2u'_2}&=&\frac{D_n(x+1)}{u_1}
+\frac{E_n(x+1)}{u_2}+\frac{F_n(x+1)}{u'_2},\nonumber\\
n&=&0,\,1,\,2.
\end{eqnarray}
Again, we encounter integrals of the types~(\ref{3t}), $v$'s being
changed by $u$'s. The terms with $u_2$ and $u_2'$ in the
denominators contribute only, while those with $u_1$ disappear in
the limit $\epsilon\to 0$, after the integration over $dx$. Using
the formulas~(\ref{apb3}) and~(\ref{apb4}) from
Appendix~\ref{appB}, we obtain after some transformations:
\begin{multline}
\label{eq16}
{\cal G}_{\rho}^{(x=1)}  =
\gamma_+\gamma_{\rho}\gamma_+\,\left\{\pi^2+2\pi i
\left(\log\frac{L}{a}-\log\frac{1}{\epsilon} \right)\right.\\
 \left.+2\pi iH_1({\bf k}_{\perp})\right\} \\
  + 2\pi ip_+\left[{\cal M}'_{10}\gamma_{\rho}\gamma_++
 \gamma_+\gamma_{\rho}{\cal M}_{10}\right]H_2({\bf k}_{\perp}),
\end{multline}
where
\begin{eqnarray*}
{\cal M}_{10}&=&-p_-\gamma_+-2{\bf k}_{\perp}\cd {\bg
\gamma}_{\perp}-2m,\\
{\cal M}'_{10}&=&-p'_-\gamma_+-2({\bf k}_{\perp}-{\bf
q}_{\perp})\cd {\bg \gamma}_{\perp}-2m,\\
H_1({\bf k}_{\perp})&=&\frac{p_--p'_-}{b-b'}+\log\frac{a}{b}\\
 &&+\frac{(a-b')(b-b')+bp'_--b'p_-}{(b-b')^2}\,\log\frac{b}{b'},
\\
H_2({\bf k}_{\perp})&=&\frac{1}{p_+(b-b')}\,\log\frac{b}{b'}.
\end{eqnarray*}

In order to find the form factors ${\cal F}^{(zm)}$ from the
vertex $\Gamma_{\rho}^{(zm)}$, it is convenient to cast the latter
in an explicitly covariant form. For this purpose we introduce the
four-vectors
\begin{eqnarray*}
r&\equiv&(r_-,r_+,{\bf r}_{\perp})=(p_-,0,-{\bf k}_{\perp}),\\
r'&\equiv&(r'_-,r'_+,{\bf r}'_{\perp})=(p'_-,0,{\bf q}_{\perp}-{\bf
k}_{\perp}).
\end{eqnarray*}
Then, taking the sum of Eqs.~(\ref{eq7}) and~(\ref{eq16}),
substituting it into Eq.~(\ref{eq4a}), and changing everywhere
$\gamma_+\to {\sla \omega}$, $p_+\to \omega\cd p$, we arrive at
the desired result:
\begin{multline}
\label{eq19}
 \Gamma_{\rho}^{(zm)}=-\frac{g^2}{32\pi^3}\int
d^2k_{\perp}\,\left\{\frac{{\sla \omega}\omega_{\rho}}{(\omega\cd
p)^2}\,H_1({\bf k}_{\perp}) \right.\\
\left.- \left[({\sla
r}'+m)\gamma_{\rho}\frac{{\sla \omega}}{\omega\cd p}+\frac{{\sla
\omega}}{\omega\cd p}\gamma_{\rho}({\sla r}+m)\right]\,H_2({\bf
k}_{\perp})\right\}.
\end{multline}
To find the form factors, one should substitute
$\Gamma_{\rho}^{(zm)}$ into Eq.~(\ref{OO}), instead of
$\Gamma_{\rho}$, then find $c_{1-5}$ by means of
Eqs.~(\ref{contrc}), and finally revert the system of linear
equations~(\ref{eqc5}) taken for $\alpha=0$. We give here the
result:
\begin{eqnarray*}
&&F_{1}^{(zm)}=F_{2}^{(zm)}=0,
\\
&&B_i^{(zm)}=-\frac{g^2}{32\pi^3}\int d^2k_{\perp}b_i({\bf
k}_{\perp}),
\end{eqnarray*}
where
\begin{eqnarray*}
\label{f1k} && b_1({\bf k}_{\perp})=H_2({\bf k}_{\perp}), \quad
b_2({\bf k}_{\perp})=-2H_2({\bf k}_{\perp}),
\\
&&b_3({\bf k}_{\perp})=\frac{1}{m^2}\left[H_1({\bf k}_{\perp})
-2({\bf k}_{\perp}\cd {\bf q}_{\perp}+m^2)H_2({\bf
k}_{\perp})\right].
\end{eqnarray*}
It is convenient to use the following representations:
\begin{eqnarray*}
H_1({\bf k}_{\perp})&=&{\displaystyle \int_0^1dz\left[
\frac{zp_-+(1-z)p'_-+a-b'}{bz+b'(1-z)}\right.}\\
& & \hphantom{\int_0^1dz\left[\right.}+{\displaystyle \left.
\frac{a-b}{az+b(1-z)}\right]},\\
H_2({\bf k}_{\perp})&=&{\displaystyle \frac{1}{p_+}\int_0^1
\frac{dz}{bz+b'(1-z)}},
\end{eqnarray*}
which can be checked by direct integration. Calculating first the
integrals over $d^2k_{\perp}$ and then over $dz$, we find the zero
mode contribution to the form factors, Eqs.~(\ref{BTFLFZM}).
%

\section{\label{appB} Calculation of  the  zero-mode integrals}
Finding the zero mode contribution to the self-energy and EMV
requires calculation of the asymptotic value of the integral
\begin{equation}
\label{apb1}
I_1=\int_{-\epsilon}^{\epsilon}dx\int_{-L}^{L}\frac{dk_-}{xk_--b-xh+i0}
\end{equation}
at $L\to\infty$ and $\epsilon\to 0$ ($b$ and $h$ are considered as
finite quantities independent both of $k_-$ and $x$). As we shall
see below, the real part of $I_1$ is finite in this limit, so that
\begin{eqnarray*}
\mbox{Re}\,I_1&=&\mbox{Re}\,\int_{-\epsilon}^{\epsilon}dx
\int_{-\infty}^{+\infty}\frac{dk_-}{xk_--b-xh+i0}\\
&=&\mbox{Im}\left[\lim_{\lambda\to
0}\int_{-\epsilon}^{\epsilon}dx\int_{-\infty}^{+\infty}dk_-\right.\\
&&\mbox{\ \ \ \ \ \ }\times\left.\int_{\lambda}^{+\infty}dy\, e^{i(xk_--b-xh+i0)y}\right]\\
&=&2\pi\mbox{Im}\left[\lim_{\lambda\to 0}
\int_{\lambda}^{+\infty}dy\,e^{-i(b-i0)y}\right.\\
&&\mbox{\ \ \ \ \ \ }\times\left.\int_{-\epsilon}^{\epsilon}dx\,
e^{-ihxy}\delta(xy)\right]\\
&=&-2\pi\int_0^{+\infty}dy\,\frac{\sin
by}{y}\\
&=&-\pi^2\mbox{sgn}(b).
\end{eqnarray*}
The imaginary part of $I_1$ is divergent in $L$ and $\epsilon$,
and we have to retain finite values of the cutoffs:
$$
\mbox{Im}\,I_1=-\pi\int_{-\epsilon}^{\epsilon}dx
\int_{-L}^{L}dk_-\,\delta(xk_--b-xh).
$$
Using the  identity  $\delta(z)=\delta(-z)$, we can easily see
that $\mbox{Im}\,I_1$ does not change under the transformation
$b\to -b$. It is thus legitimate to substitute $b$ by $|b|$.
Calculating the integral by means of the  $\delta$-function, we
get
\begin{eqnarray*}
\mbox{Im}\,I_1&=&-\pi\left(\int_{-\epsilon}^{-\frac{|b|}{L+h}}\frac{dx}{|x|}
+\int_{\frac{|b|}{L-h}}^{\epsilon}\frac{dx}{x}\right)\\
&=&
-\pi\left(\log\frac{L^2-h^2}{b^2}+2\log\epsilon\right).
\end{eqnarray*}
Collecting the results together and neglecting the terms of order
 $(1/L)^2$ and higher,  we finally obtain
\begin{equation}
\label{intzm1}
I_1=-\pi^2\mbox{sgn}(b)-2\pi
i\left(\log\frac{L}{|b|}-\log\frac{1}{\epsilon}\right).
\end{equation}

Let us now  consider  the integral needed for the calculation of
the zero-mode contribution to the EMV:
\begin{equation}
\label{apb2}
I_2=\int_{-\epsilon}^{\epsilon}\frac{dx}{x}\int_{-L}^{L}
\frac{dk_-}{xk_--b-xh+i0},
\end{equation}
where the principal value prescription for the pole at $x=0$ is
implied. The real part of $I_2$ is found by direct integration:
\begin{eqnarray*}
\mbox{Re}\,I_2&=&\int_{-\epsilon}^{\epsilon}\frac{dx}{x}
\left(\mbox{P.V.}\int_{-L}^{L}\frac{dk_-}{xk_--b-xh}
\right)\\
&=&\int_{-\epsilon}^{\epsilon}\frac{dx}{x^2}\,\log\left|
\frac{xL-b-xh}{xL+b+xh}\right|\\
&=&\frac{4h}{\epsilon
L}+O\left[\left(\frac{h}{\epsilon L}\right)^3\right].
\end{eqnarray*}
We  have expanded  the result in a series in powers of
$1/(\epsilon L)$, retaining the leading term only. Note that
according to our prescription we should first tend $L$ to
infinity, while keeping $\epsilon$ finite, and then tend
$\epsilon$ to zero. Under such a  convention,  we obtain
$\mbox{Re}\,I_2=0$. The imaginary part of $I_2$ reads
\begin{eqnarray*}
\mbox{Im}\,I_2&=&-\pi\int_{-\epsilon}^{\epsilon}\frac{dx}{x}
\int_{-L}^{L}dk_-\delta(xk_--b-xh)\\
&=&
-\pi\int_{-\epsilon}^{\epsilon}\frac{dx}{x^2}\int_{-xL-b-xh}^{xL-b-xh}dy
\delta(y)\\
&=&\frac{2\pi h}{b}.
\end{eqnarray*}
Finally,
$$
I_2=\frac{2\pi i h}{b}.
$$

From the results obtained above it is easy to derive the following
expressions for the generalized integrals
\begin{eqnarray}
\tilde{I}_1&=&\int_{-\epsilon}^{\epsilon}dx\,f(x)\int_{-L}^{L}
\frac{dk_-}{xk_--b-xh+i0}\label{apb3} \\
&=&-\left[\pi^2\mbox{sgn}(b)+2\pi
i\left(\log\frac{L}{|b|}-\log\frac{1}{\epsilon}\right)\right]f(0),\nonumber\\
\tilde{I}_2&=&\int_{-\epsilon}^{\epsilon}dx\,\frac{f(x)}{x}\int_{-L}^{L}
\frac{dk_-}{xk_--b-xh+i0}\label{apb4} \\
&=&\frac{2\pi i h}{b}f(0)\nonumber\\
&&-\left[\pi^2\mbox{sgn}(b)+2\pi
i\left(\log\frac{L}{|b|}-\log\frac{1}{\epsilon}\right)\right]f'(0),\nonumber
\end{eqnarray}
where  $f(x)$ is an arbitrary function  supposed to be
smooth and finite at $x=0$.
%


\begin{thebibliography}{10}
\bibitem{cdkm}
J.~Carbonell, B.~Desplanques, V.A.~Karmanov and J.-F.~Mathiot,
Phys. Rep. {\bf 300}, 215 (1998).

\bibitem{bpp}
S.J.~Brodsky, H.C.~Pauli, and S.S.~Pinsky,
Phys. Rep.  {\bf 301}, 299 (1998).

\bibitem{Br_et_al_03}
S.J.~Brodsky, J.R.~Hiller, G.~McCartor, Annals Phys. {\bf 305}
 266 (2003).

\bibitem{Br_et_al_04}
S.J.~Brodsky, V.A.~Franke, J.R.~Hiller, G.~McCartor, S.A.~Paston,
E.V.~Prokhvatilov, Nucl. Phys. {\bf B 703}, 333 (2004).

\bibitem{Br_et_al_06}
S.J.~Brodsky, J.R.~Hiller, G.~McCartor, Annals Phys. {\bf
321}, 1240 (2006).

\bibitem{BCKM_02}
D.~Bernard, Th.~Cousin, V.A.~Karmanov, J.-F.~Mathiot, Phys. Rev.
{\bf D 65}, 025016 (2002).

\bibitem{kms_04}
V.A.~Karmanov, J.-F.~Mathiot, A.V.~Smirnov, Phys. Rev. {\bf D 69},
045009 (2004).

\bibitem{mks_05}
J.-F.~Mathiot, V.A.~Karmanov, A.V.~Smirnov, Nucl. Phys. {\bf B}
(Proc. Suppl.) {\bf 161}, 160 (2006).

\bibitem{lb95}
N.E.~Ligterink and B.L.G.~Bakker, Phys. Rev. {\bf D 52}, 5917,
5954 (1995).

\bibitem{Nico1}
N.C.J.~Schoonderwoerd and B.L.G.~Bakker, Phys. Rev. {\bf D 58},
 025013 (1998).

\bibitem{frederico}
J.P.B.C.~de~Melo and T.~Frederico, Phys. Rev. {\bf C 55}, 2043
(1997);
 J.P.B.C.~de~Melo, J.H.O.~Sales, T.~Frederico and
P.U.~Sauer, Nucl. Phys. {\bf A 631}, 584c (1998);
 J.P.B.C.~de~Melo, T.~Frederico, H.W.L.~Naus and P.U.~Sauer,
 Nucl. Phys. {\bf A 660}, 219 (1999);
 H.W.L.~Naus, J.P.B.C.~de~Melo and T.~Frederico, Few-body Systems,
 {\bf 24}, 99 (1998);
 J.P.B.C.~de~Melo and T.~Frederico, Braz.J.Phys. {\bf 34}, 881,
 (2004).

\bibitem{bakker_02}
B.L.G.~Bakker, H.-M.~Choi  and Ch.-R.~Ji, Phys. Rev. {\bf D 65},
116001 (2002).

\bibitem{miller_03}
B.C.~Tiburzi and G.A.~ Miller, Phys. Rev. {\bf D 67}, 054014,
054015 (2003).

\bibitem{bissey}
F.~Bissey and J.-F.~Mathiot, Eur. Phys. J.  {\bf C 16}, 131
(2000).

\bibitem{jaus_2003}
W.~Jaus, Phys. Rev. {\bf D 67}, 094010 (2003).

\bibitem{bakker_05}
B.L.G.~Bakker, M.A.~DeWitt, Ch.-R.~Ji and Yu.~Mishchenko, Phys.
Rev. {\bf D 72}, 076005 (2005).

\bibitem{jibak1}
Ch.-R.~Ji, B.L.G.~Bakker and H.-M.~Choi, Nucl. Phys. {\bf B}
(Proc. Suppl.) {\bf 161}, 102 (2006).

\bibitem{dugne}
J.-J.~Dugne, V.A.~Karmanov and J.-F.~Mathiot, Eur. Phys. J. {\bf
C22}, 105 (2001).

\bibitem{karm76}
V.A.~Karmanov, ZhETF, {\bf 71}, 399 (1976) [transl.: JETP, {\bf
44}, 210 (1976)].

\bibitem{km96}
V.A.~Karmanov and J.-F.~Mathiot, Nucl. Phys. {\bf A 602}, 338
(1996).

\bibitem{Pierre}
P.~Grang\'e and E.~Werner, Nucl. Phys. {\bf B} (Proc. Suppl.) {\bf
161}, 75 (2006).

\end{thebibliography}
\end{document}